\documentclass[prb,showpacs,twocolumn,floatfix,superscriptaddress]{revtex4}
\usepackage{hyperref,fontenc,times,amsmath,amsfonts,amssymb,graphics,graphicx,epsfig,color,bbm,bm,units,mathtools,hhline}

\newcommand{\dd}{\mathrm{d}}
\newcommand{\ee}{\mathrm{e}}

\DeclareMathOperator{\Tr}{Tr}
\DeclareMathOperator{\ad}{ad}

\newlength{\depthofsumsign}
\newcommand{\nsum}[1][1.4]{
	\mathop{%
		\raisebox
		{-#1\depthofsumsign+1\depthofsumsign}
		{\scalebox
			{#1}
			{$\displaystyle\sum$}%
		}
	}
}

\DeclarePairedDelimiterX\braket[2]{\langle}{\rangle}{#1\delimsize\vert}
\DeclarePairedDelimiterX\braxket[3]{\langle}{\rangle}{%
  #1\,\delimsize\vert\,#2\,\delimsize\vert\,#3}

\makeatletter
\renewcommand*\env@cases[1][1.2]{%
  \let\@ifnextchar\new@ifnextchar
  \left\lbrace
  \def\arraystretch{#1}%
  \array{@{}l@{\quad}l@{}}%
}
\makeatother

\begin{document}

\title{`Goldilocks' Probes for Noisy Interferometry via Quantum Annealing to Criticality.}

\author{Gabriel A. Durkin} \email{Gabriel.Durkin@qubit.org}
\affiliation{Berkeley Quantum Information and Computation Center, University of California, Berkeley, CA 94720}
\affiliation{Peliquan Technologies, 950 Franklin St., Suite 1, San Francisco, CA 94109}

\date{\today}

\begin{abstract}
Quantum annealing is explored as a resource for quantum information beyond solution of classical combinatorial problems. Envisaged as a generator of robust interferometric probes, we examine a Hamiltonian of $N>> 1$ uniformly-coupled spins subject to a transverse magnetic field. The discrete many-body problem is mapped onto dynamics of a single one-dimensional particle in a continuous potential. This reveals all the qualitative features of the ground state beyond typical mean-field or large classical spin models. It illustrates explicitly a graceful warping from an entangled unimodal to bi-modal ground state in the phase transition region. The transitional `Goldilocks' probe has a component distribution of width $N^{2/3}$ and  exhibits characteristics for enhanced phase estimation in a decoherent environment.  In the presence of realistic local noise and collective dephasing, we find this probe state asymptotically saturates ultimate precision bounds calculated previously. By reducing the transverse field adiabatically, the Goldilocks probe is prepared in advance of the minimum gap bottleneck, allowing the annealing schedule to be terminated `early'. Adiabatic time complexity of probe preparation is shown to be linear in $N$. 
\end{abstract}

\maketitle

\section{Introduction}

In quantum metrology\cite{demkowicz2015quantum,giovannetti2011advances} we often seek to estimate a continuous time-like parameter associated with unitary evolution. Even without a direct Hermitian observable for time or phase, one can determine bounds on the mean-squared error of estimated values as a function of $N$, the number of qubits, particles, spins or photons involved in the measurement.  The lowest bounds are associated with initializing the instrument in a particular entangled quantum configuration of the $N$ qubits, known as a `probe' state. Without entanglement, the performance cannot exceed the precision resulting from sending qubits through the instrument one at a time. Large spin and mean-field models used to describe many-body systems typically ignore entanglement altogether. Curiously, in a noisy setting the most entangled states do not offer the greatest precision \cite{huelga_improvement_1997}.

In the noiseless case, it has been known for some time that the optimal configuration of the $N$ qubits is the NOON\cite{sanders_noon,dowling_quantum_2008} or GHZ \cite{greenberger1989going} state. This is an equal superposition of the two extremal eigenstates of the phase-encoding Hamiltonian. Subsequently, however, we have come to understand that this state offers sub-optimal performance in the presence of realistic noise or decoherence, and recent work has unveiled a new family of optimal probe states for noisy metrology \cite{knysh2014true}.   

Unfortunately, this result brings with it the new challenge of generating such probes. The asymptotic analysis that uncovered the optimal states indicates also that, for any large-$N$ probes, there will be a precision penalty for those with  discontinuities in the distribution of components. (This is the case with the NOON/GHZ state.). 

For a spin Hamiltonian like $\hat{J}_z$ associated with phase or frequency estimation, optimal probes typically inhabit the fully-symmetric subspace of largest overall spin $j = N/2$. For $N \gg 1$ optimality is achieved by a smooth unimodal distribution of amplitudes, a ground state for a one-dimensional particle trapped between two repulsive Coulomb sources\cite{knysh2014true}.   The optimal distribution width $\Delta \hat{J}_z$ is dependent on the noise strength and is typically wider than $\sqrt{N}$, i.e. it is anti-squeezed in the $z$-direction. Had the optimal probe  such a `square-root' width, it would be easily produced by rotating an $N$ -spin coherent state by $\pi/2$ around the $x$-axis via an optical pulse. This state is a simple product state of the component spins --  creating `wider' optimal probes introduces partial, or `just the right amount' of entanglement to the ensemble.

In this paper, we explore techniques to generate such quantum probes, balancing sensitivity against robustness.

\begin{figure}
	\includegraphics[width=3.3in]{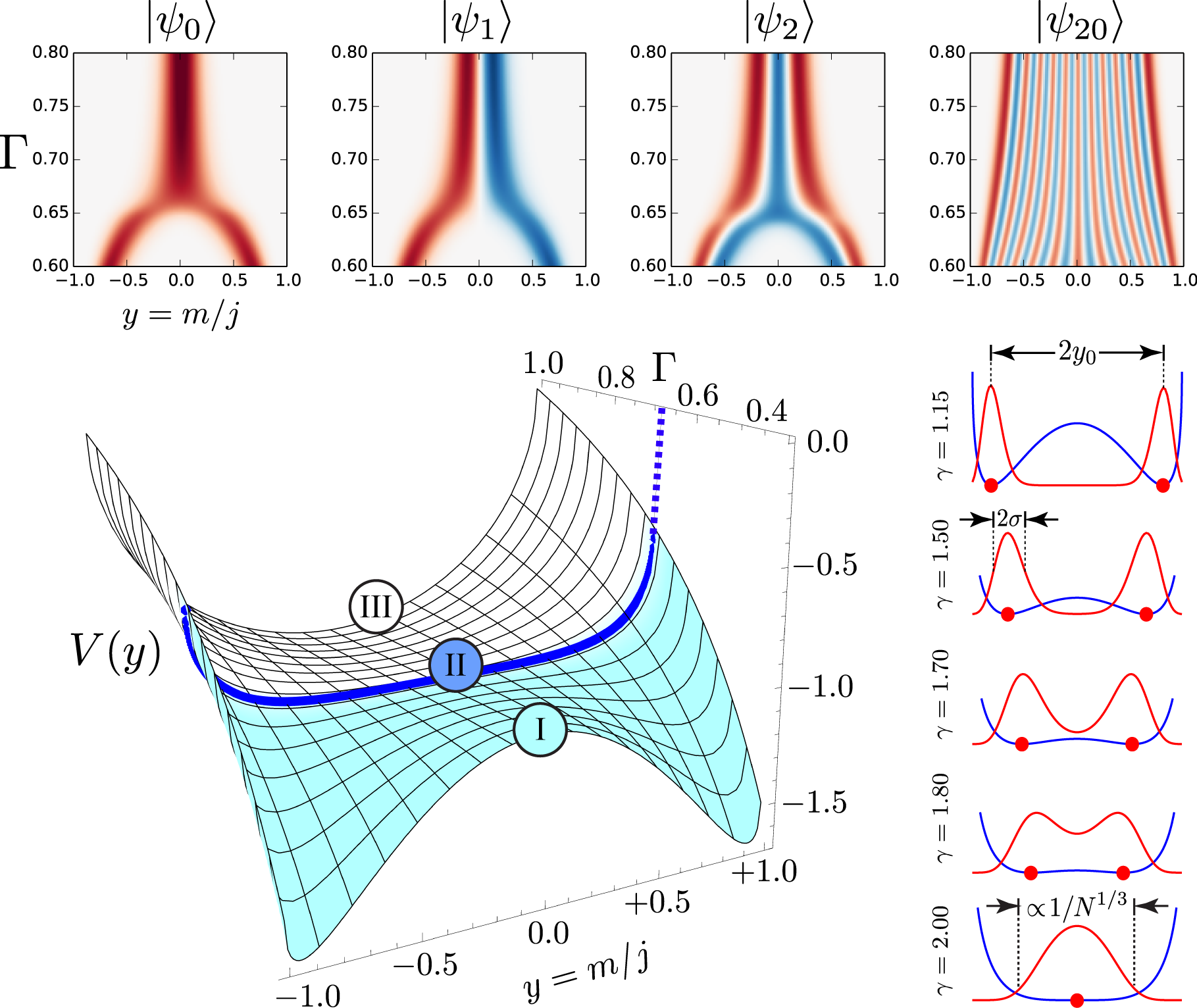}
	\caption{\label{pseudo}
		\textbf{Pseudo-potential} $V(y)$ from eqn.\eqref{Vy} that results in the large $N$ limit from mapping the quadratic spin Hamltonian onto a 1-D particle in a potential well. The evolution from a single well in Region III  through a critical Region II to a double-well in Region I is apparent as the transverse field is decreased, or equivalently, as the annealing parameter $\Gamma$ is swept from $1$ to $0$. In Region III where the transverse field is strongest, the ground state will be all spins aligned with the field along the x-axis. In the variable $y$, this is a  Gaussian state of width $1/\sqrt{N}$. Gaussian states are the ground state of a quadratic potential. At $\Gamma = 1$, however, the pseudo-potential above looks like a semi-circle. There is effectively no disparity because the ground state is very narrow for large N, and has very low probability away from $y= 0$; within this locale $V(y)$ also looks quadratic. The dark blue contour in Region II indicates the  quartic form of the potential at the critical point, where the annealing parameter $\Gamma_c = 2/3$ or $\gamma_c = \Gamma_c/(1-\Gamma_c) = 2 $. For $N=100$, the right-hand-side vertical sequence shows the ground state of the original spin Hamiltonian (in red)  overlaid by the pseudo-potential  (in blue), with potential minima indicated by red dots.  For an $N=200$ ensemble, eigenstates $| \psi_n \rangle$ with $n \in \{0,1,2,20 \}$  pass through the critical region in the upper four panels; lower energy states exhibit a \emph{pitchfork bifurcation} associated with the phase transition. (Positive amplitudes in red, negative in blue.)}
\end{figure}

\section{Hamiltonian for Probe Preparation}
Bearing in mind the ideal characteristics above, one might start to imagine how such broad, smooth, unimodal probe-state distributions could be engineered. To this end, one of the simplest non-trivial quantum systems that can be investigated is one with an equal $\hat{\sigma}_{z}^{(1)} \hat{\sigma}_{z}^{(2)}$ coupling between all pairs of qubits in the presence of a transverse field. The field strength increases monotonically with an `annealing' parameter $\Gamma$:
\begin{equation} \label{spinH}
\hat{H} = - \Gamma \frac{\hat{J}_x}{j} -  (1- \Gamma) \frac{\hat{J}_z^2}{j^2} 
\end{equation}
where $2 \hat{J}_{z} =   \hat{\sigma}_{z}^{(1)}+\hat{\sigma}_{z}^{(2)}+ \hat{\sigma}_{z}^{(3)} + \dots$ In this scaled form, $| \langle \hat{H} \rangle| \leq 1$; so $j \hat{H}$ corresponds to actual energies. This system exhibits a continuous quantum phase transition, as follows. Initializing the system in the ground state of a strong transverse field  $\Gamma \lesssim 1$ , all spins are aligned with the $x$-axis (this is the coherent spin state discussed in the introduction). Then, as the field is gradually attenuated, the parameter $\Gamma$ decreases to a critical value $\Gamma_c = 2/3$, at which point the ground state warps \emph{continuously} into a qualitatively different bimodal NOON-like profile (exactly a NOON state when $\Gamma =0$).  If the annealing proceeds slowly enough, the spins will remain in the instantaneous ground state at all times; this is adiabatic passage. How realistic are such annealing dynamics? Quadratic terms like $\hat{J}_z^2$ appear frequently in models of two-mode Bose-Einstein condensates\cite{milburn1997quantum,cirac1998quantum} (BEC) , describing collisional processes. The two modes may correspond to a single condensate in a double-well potential, or a mixture of atoms in two distinct hyperfine states in a single potential. One of the earliest proposals for generating the spin-spin couplings was introduced in the context of ion traps illuminated by two laser fields\cite{molmer1999multiparticle}. This Hamiltonian is also referred to as the isotropic \emph{Lipkin-Meshkov-Glick} (LMG) model \cite{orus2008equivalence}, an infininte-range Ising model with uniform couplings. The LMG model can provide an effective description of quantum gases with long range interactions\cite{baumann2010dicke}. In terms of metrology, the precision offered by some Ising models in a decoherence-free setting was given careful examination recently in Ref.\onlinecite{skotionitis2015quantum}. 

 Similarities exist with the dynamics of of a single-mode oscillator (optical field in a cavity), coupled adiabatically to a collection of spins or atoms via the Dicke Hamiltonian \cite{liberti2010finite}, where the coherence length of the field is much larger than the physical extent of the particle ensemble. The single-mode field introduces an effective ferromagnetic spin-spin coupling.  The intensity of light emitted into the Dicke super-radiant phase can also be utilized for high-precision thermometry \cite{gammelmark2011phase} for probes prepared in close proximity to the critical point. 

It is not our goal to measure the temperature, evolution time, transverse field or any other `native' property  of this system. Rather, the objective is to utilize the annealing dynamics as a resource for engineering robust high-precision probes for interferometry in noisy environments.

The Hamiltonian of eqn.\eqref{spinH} has been considered previously for interferometry in works that suggest that it is a good source of squeezing and, as such, should lead to better precision. Those prior works\cite{law2001coherent,rojo2003optimally}, however, did not describe the dynamics through the critical region, where we believe maximum precision is possible. We now know\cite{knysh2014true} that the extent of probe squeezing is not a good quantification of precision in a noisy interferometer (the most squeezed input states may be some of the most fragile).

\section{Simple Mapping onto a particle in a 1-D continuous potential}

To capture fully the behaviour of this discrete spin system at large $N$ throughout the phase transition (and determine if it has appropriate properties for noisy interferometry), we map it onto a continuous particle problem.

The idea of mapping a quadratic spin Hamiltonian in a transverse field onto a one-dimensional particle in a potential is not a new one; many examples exist in the literature \cite{scharf1987tunnelling,garanin1998quantum,ocak2003effective,ulyanov1992new,sciolla2011dynamical}. A typical approach involves `bosonization' of the spin operators into combinations of $\hat{a}$ and $\hat{a}^\dagger$, using either Holstein-Primakoff \cite{holstein1940field} or Villain \cite{villain1974quantum} transformations. Then, after identifying quadratures $\hat{x}= (\hat{a} + \hat{a}^\dagger)/2 $ and $\hat{p}= -i (\hat{a} - \hat{a}^\dagger)/2$,  an operator differential equation in $x$ and $\hat{p} = -i d/d x$ is produced that, after some approximations, may resemble a Schr\"odinger equation. (One may choose to linearize the boson operators about the mean-field direction.)

To understand the behaviour near the critical point, the mapping must remain faithful to the original discrete spin dynamics, both qualitatively and quantitatively (to leading order, when finite size effects are considered). As a caveat, it is admitted that certain subtle phenomena may not be captured, e.g. exponentially small ground-state splitting that occurs for a weak transverse field. This requires precision calculation of small probability `tails' deep inside the barrier dividing a double potential well \cite{garg2000tunnel}, e.g. region I of FIG.\ref{pseudo}. Luckily, interferometric precision is quantified largely by the bulk probabilities of the probe-state concentrated at the bottom of the potential wells; any evanescent amplitude in the forbidden region makes an exponentially subordinate contribution.

\begin{figure}[h!]
	\includegraphics[width=3.4in]{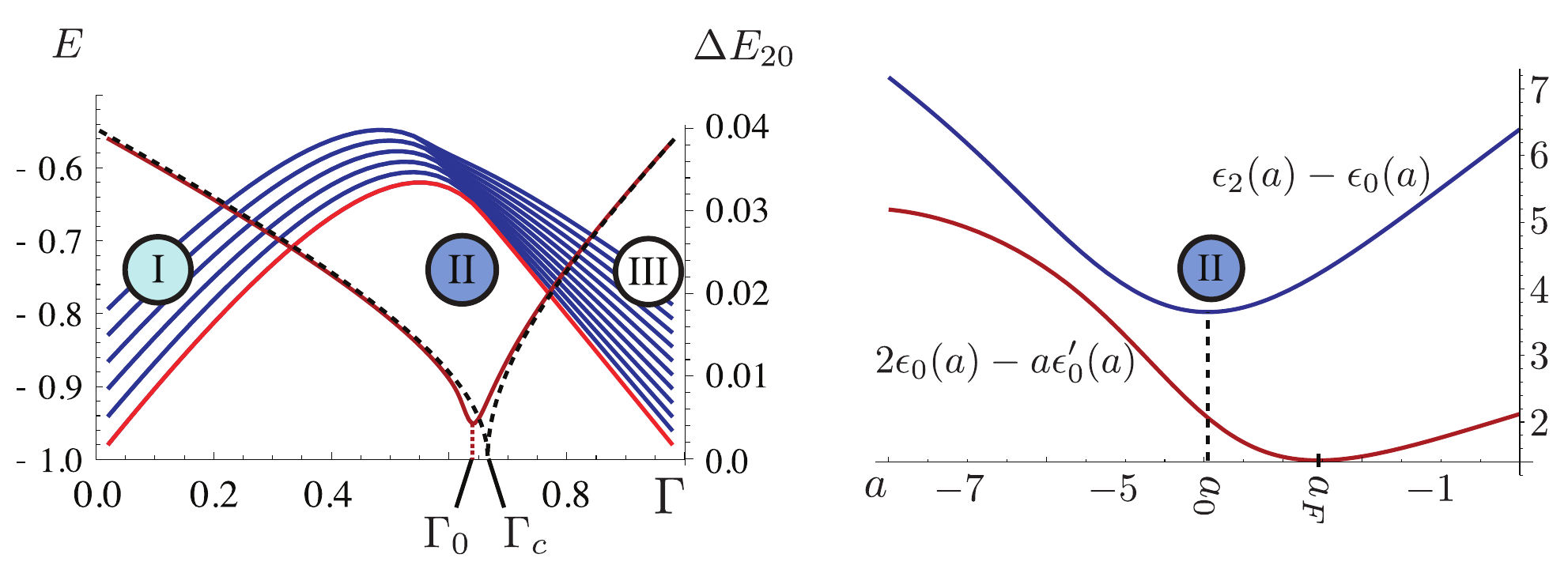}
	\caption{  \label{wavy}
		\emph{(left plot)} \textbf{Energy Levels} $E_n$ for $n \in [0,11]$ (blue curves, left vertical axis) and  ground-state energy gap $\Delta E_{20} = E_2 -E_0$ (dark red curve, right vertical axes) for an $N=100$ spin ensemble as the annealing parameter $\Gamma$ is swept from $0$ to $1$ through the location of the minimum gap $\Gamma_0$ and the critical $\Gamma_c$ where the phase transition occurs in the thermodynamic limit, $N \gg 1$.  In Regions I and III,  energy levels are uniformly distributed $\Delta E \propto 1/j$, although in Region I neighbouring even and odd numbered levels pair up; only exponentially small gaps separate them. Dashed curve  indicates the energy gap in the thermodynamic limit of eqn.\eqref{omegat}. \emph{(right plot)}	\textbf{Region II Landmarks:} Numerical results in the scale-free system are shown for the precision penalty factor (red curve) from eqn.\eqref{penalty}, and ground state gap (blue curve) near the critical point $a=0$  (or $\Gamma = 2/3$). It is observed that $a_F$, the point of maximum precision is to the right of the minimum gap $a_0$, and therefore $\Gamma_F > \Gamma_0$, also. Any annealing schedule from strong to weak transverse field will reach the point of highest precision before it encounters the minimum gap.}
\end{figure}

Using notation $\hat{J}_z | m \rangle = m | m \rangle$ for eigenstates of $\hat{J}_z$ labelled by magnetic quantum numbers $m \in \{ -j, -j+1, \dots, +j\}$ one may represent the quantum ground state as a vector of amplitudes $\psi_m$ in the $| m \rangle$ basis, 
\begin{equation}
| \psi_0 \rangle = \sum_{m = -j}^{+j} \psi_m | m \rangle
\end{equation}
Take the overlap $\langle m | \hat{H}| \psi_0 \rangle$ in the eigen-equation as
\begin{equation}
- \langle m | \left\{ \Gamma \frac{\hat{J}_x}{j} + (1- \Gamma) \frac{\hat{J}_z^2}{j^2}  \right\} | \psi_0 \rangle = E_0 \langle m | \psi_0 \rangle \; .
\end{equation}
Remembering $\langle m | \psi_0 \rangle = \psi_m$ and the definition in terms of ladder operators, $\hat{J}_x = (\hat{J}^{(+)}+\hat{J}^{(-)})/2$ where  $\hat{J}^{(\pm)} | m \rangle = \sqrt{j^2 - m^2 +j \mp m} | m \pm 1 \rangle$, some book-keeping produces:
\begin{align}
\psi_{m-1} \sqrt{1+ \frac{1}{j-m}} + \psi_{m+1} \sqrt{1+ \frac{1}{j+m}}   =  \nonumber \\ - \frac{2j}{\gamma} \frac{1}{\sqrt{j^2-m^2}} \left( \frac{E_0}{1- \Gamma} + \frac{m^2}{j^2}\right) \psi_m
\end{align}
where $\gamma = \Gamma / (1 -\Gamma)$ is the annealing `ratio'.
Now, assuming $j \gg 1$ one can transform into a continuous variable picture, effectively the reverse technique to solving differential equations numerically by discretizing variables. We assume a small parameter $\delta = 1/j$ for asymptotic expansions, and introduce a continuous variable $y = m/j \in  [-1,1]$,  mapping $\psi_m \mapsto \psi(y)$ and $\psi_{m \pm 1} \mapsto \psi(y \pm \delta)$. Also, assuming features change smoothly on a scale $\sim \delta$ one may define derivatives:
\begin{subequations}
\begin{align}
\frac{\psi(y + \delta) - \psi(y - \delta)}{2 \delta} &\mapsto \frac{d \psi}{d y} \label{grad-def} \\
\frac{\psi(y + \delta) + \psi(y - \delta) - 2 \psi(y)}{\delta^2} &\mapsto \frac{d^2 \psi}{d y^2}
\end{align}
\end{subequations}
Having transformed from a difference equation to a differential equation, the eigen-equation becomes a Schr\"odinger equation for a one-dimensional particle of variable mass in a pseudo-potential, as follows:
\begin{equation}\label{schrod1d}
 \left[ \frac{1}{2} \hat{P} \hat{M}^{-1} \hat{P} -  \hat{M}^{-1}  -\frac{y^2}{\gamma} \right] \psi(y) = \frac{E_0}{\Gamma} \psi(y) \: ,
\end{equation}
given an inverse mass operator, $\hat{M}^{-1}(y) = \sqrt{1-y^2} + \frac{\delta}{2} \frac{1}{\sqrt{1-y^2}}$, and a momentum operator $\hat{P} = - i \delta \frac{d}{d y}$. When solved numerically, the eigenstates of this continuous differential equation map faithfully onto the probability amplitudes for the original quadratic spin problem. See FIG.\ref{greenred} in the appendix. Variable-mass Schrodinger equations have been tackled analytically previously, e.g. in Refs.\onlinecite{alhaidari2002solutions,jha2011analytical}.

 Written as $ \hat{P} \hat{M}^{-1} \hat{P}/2$ the kinetic energy operator takes the form of a manifestly Hermitian operator. The pseudo-potential is 
\begin{equation} \label{Vy}
V(y)  = - \frac{y^2}{\gamma} - \sqrt{1-y^2} - \frac{\delta}{2} \frac{1}{\sqrt{1-y^2}} \; .
\end{equation}
This potential is depicted for $\delta \ll 1$, i.e. $N \gg 1$, in FIG.\ref{pseudo}, taking the form of either a single or double well. For large $N$, the distribution will be concentrated at the bottom of these wells at $y_0$ (red dots in FIG.\ref{pseudo}), and one may make the simplification in the kinetic term: $\hat{M}^{-1}(y) \mapsto 1/M(y_0) = 1/M_\gamma$ where $y_0$ will be a function of the parameter $\gamma$.

As such, when $\delta \ll 1$  the Hamiltonian becomes
\begin{equation}
\hat{H} \mapsto \Gamma \left( \frac{\hat{P}^2}{2 M_{\gamma}}  - \frac{y^2}{\gamma} - \sqrt{1-y^2} \right) \; .
\end{equation}

%

\section{Characteristic Energy and Length scales at Critical Point} \label{critical-scaling}
Our hope is that near criticality, the ground state may have properties that make it a promising candidate for noisy interferometry. Interestingly, the potential terms quadratic in $y$ can be made to vanish at a critical transverse field $\Gamma_c$. For states strongly concentrated near $y=0$,  expand $V(y)$ as a Taylor series.
\begin{equation} \label{pseudopot}
V(y) \approx - 1+\frac{y^2}{2} \left(1 - \frac{2}{\gamma } \right)+ \frac{y^4}{8} \: .
\end{equation}
It is seen that the leading order term of $V(y) \approx y^4/8$ near $\gamma_c =2$, ($\Gamma_c = 2/3$), when $\delta \ll 1$. One might expect the quartic ground state also to have a distribution of width scaling greater than $\sqrt{N}$ and, as such, may be a robust probe in noisy conditions. The width can be checked by employing a Symanzik scaling argument.  (A similar approach was used in Ref.\onlinecite{liberti2010finite} to recover finite size corrections to the critical exponents at exactly $\gamma_c =2$.) Due to the reflection symmetry about $y=0$ the inverse mass $1/M_\gamma$ has a minimum value $1/M_2$ there and its first derivatives vanish. Again, expand $V(y)$ to the fourth power and the Schr\"odinger equation in the vicinity of $y=0$ becomes
\begin{equation}
\left[-\frac{d^2}{dy^2} + \nu y^2 +g y^4\right] \psi_n = \tilde{E}_n \psi_n \label{reduced-schrod}
\end{equation}
where $\tilde{E}_n = \left(\frac{E_n M_2}{\Gamma}+1 \right) \frac{2 }{\delta^2}$ , $\nu = 4 g \left( 1 - \frac{2}{\gamma}\right) $ and $g = \frac{M_2}{4 \delta^2} \sim N^2/16$. Now one may rewrite everything in a scale-free way in terms of a single parameter, `$a$':
\begin{equation}
\left[-\frac{d^2}{dz^2} + a z^2 + z^4\right] \phi_n = \epsilon_n (a) \phi_n \; , \label{scale-free}
\end{equation}
with scale-free coordinates $z = y g^{1/6}$, $\psi(y) =g^{1/12} \phi(z)$, $a = \nu/g^{2/3} = 4 g^{1/3}  ( 1 - 2/\gamma) $ and $\epsilon_n = \tilde{E}_n/g^{1/3}$. At the critical point $\nu = a = 0$, and the eigenvalue problem is reduced to that of the pure quartic potential. Note that the energy spectrum, including $\epsilon_0$, and the half-width of its ground-state, let's call it $z_0$, are pure numerical values. Scaling back from $\epsilon_0$ to $E_0$ indicates that the spectrum is compressed near the critical point,
\begin{equation} \label{gap}
\Delta E_{k \ell} = \frac{\Gamma  \Delta \epsilon_{k \ell} }{2 (4^{1/3}) j^{4/3}} \; \; \; \; \;  (\Gamma \approx 2/3) \; .
\end{equation} 
Remembering that $\Delta \epsilon_{k \ell}$ is a pure number, the energy gap in the original problem is compressed by $j^{-1/3}$ compared with the strong or very weak transverse field Regions I and III, where it is uniform in $j \hat{H}$, as we shall see in section \ref{GSregionI}. The compression of eigenvalues can be inspected in  FIG.\ref{wavy}  for an $N=100$ ensemble. Establishing the true length-scale for $y$ involves dividing $z_0/ g^{1/6} \propto \delta^{1/3}$ or $j^{-1/3}$, without having to recover any features of the wavefunction explicitly. Recall that $y = m/j$ the width scales as $j^{2/3}$ in $m$, or indeed $N^{2/3}$. As we hoped, this partially entangled `Goldilocks' state at the critical point has greater width than the Gaussian separable distribution of width $\sqrt{N}$ associated with a spin-coherent state (such as the ground state at $\Gamma = 1$).

\section{Location of Minimum Gap} \label{mingapdisc}
From the beginning, our desire has been to prepare a Goldilocks probe via quantum annealing -- we  reduce the transverse field adiabatically, keeping the system in the instantaneous ground state at all times. The annealing must proceed especially slowly when the gap between ground and excited states is smallest, avoiding diabatic passage into another eigenstate. It is necessary, therefore, to establish the size and location of the minimum gap during the  schedule, as this will be the dominant bottleneck affecting efficient probe preparation.

\begin{figure}[h!]
	\includegraphics[width=2.4in]{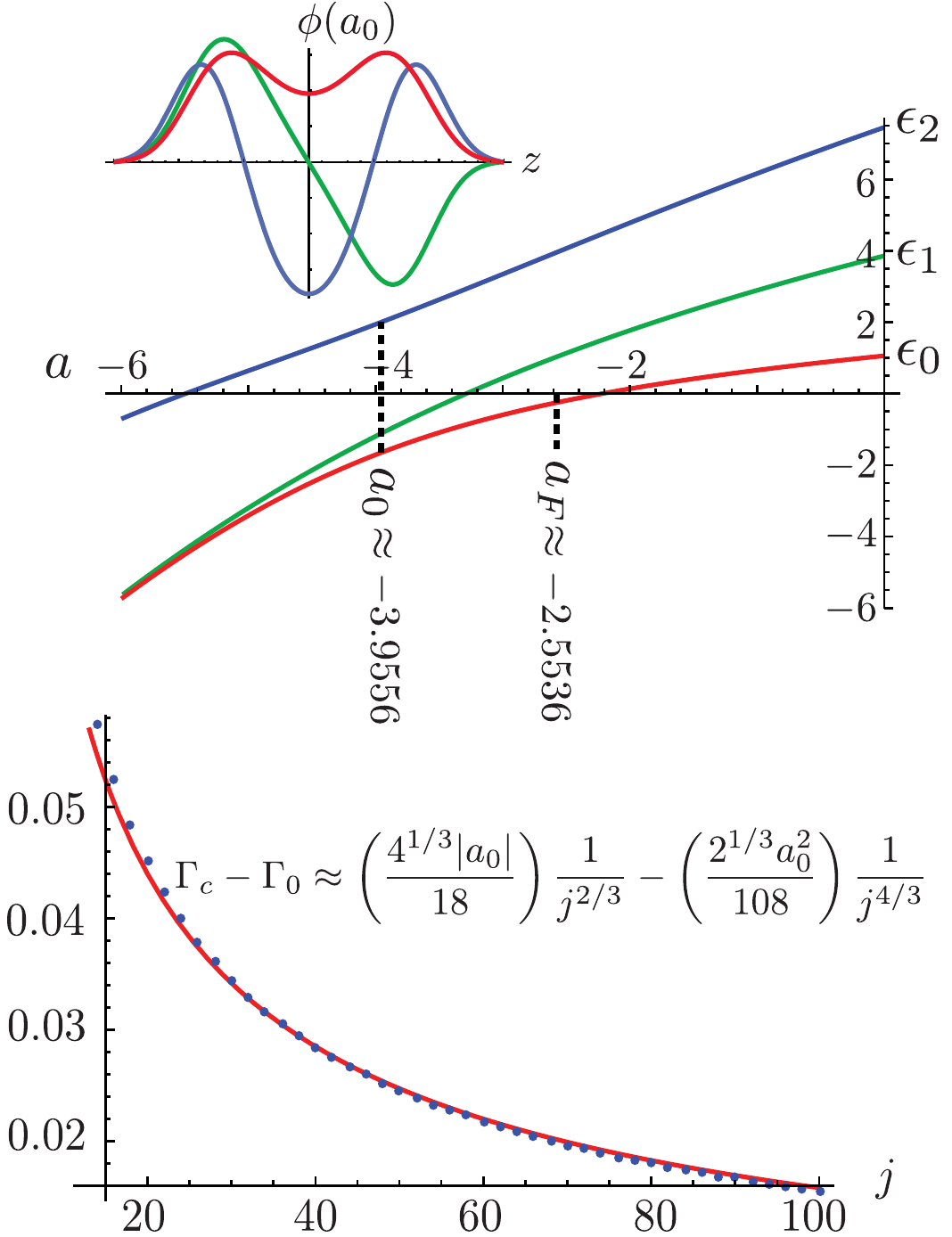}
	\caption{\label{locmin}
		\emph{(upper)} \textbf{Rescaled Spectrum:} Eigenstates $\phi_{0,1,2}$ and energy eigenvalues $\epsilon_{0,1,2}$ found numerically for the scale-free quartic potential and single parameter $a$. The minimum gap $a_0$ and  maximum precision point $a_F$ are indicated. \emph{(lower)} \textbf{Minimum Gap:} Blue points  show the difference $\Delta \Gamma$ between the value of the annealing parameter at the minimum gap $\Gamma_0$ and the critical point $\Gamma_c = 2/3$ in the original discrete spin Hamiltonian. The red curve has function $0.349/ j^{2/3}-0.183/j^{4/3}$ derived in section \ref{mingapdisc}.}
\end{figure}

Close to the critical point in the thermodynamic limit, $a \ll 1$. In the scale-free setting, one could potentially treat the $\langle z^2  \rangle$ term of eqn. \eqref{scale-free} perturbatively, $\epsilon_0(a) = \epsilon_0 (0) + a \epsilon'_0 (0) + \dots$.It turns out, however, from an exact numerical analysis  shown in FIG.\ref{wavy} (right side) and FIG.\ref{locmin} (upper plot),  that the minimum gap $\epsilon_2 (a_0) - \epsilon_0 (a_0)$ does not correspond to a convergent perturbative regime, as $| a_0 | >1$. We focus on the $0 \leftrightarrow 2$ transitions because there is no matrix element between the ground and first excited states; they have opposite parity.
%

The annealing parameter $\Gamma_0$ at the minimum gap may be identified as:
\begin{equation}
\Gamma_0 = \frac{\Gamma_c}{1-\frac{a_0}{12 g^{1/3}}} \approx \Gamma_c + \left(\frac{ 4^{1/3} a_0}{18 }  \right) \frac{1}{j^{2/3}} +\left(\frac{2^{1/3} a_0^2}{108}  \right) \frac{1}{j^{4/3}} \: , 
\end{equation} 
again recalling the relation $a= 4 g^{1/3}  ( 3 - 2/\Gamma)$. Using the numerical result $a_0 \approx - 3.9556$, the prefactor to the dominant $j^{-2/3}$ scaling of $\Delta \Gamma_c = \Gamma_c - \Gamma_0$ is approximately $0.349$. Prefactors and scaling for the leading terms are confirmed by comparison with the minimum gap of the original spin problem for different ensemble sizes $j$ in FIG.\ref{locmin} (bottom). Throughout this paper our goal is to map out this spin system's features in the Goldilocks critical region at finite ensemble size $N$. The apparent discontinuities occurring in the thermodynamic limit $N \sim  \infty$ provides no guidance here, as the Goldilocks zone vanishes in this limit.

From the shape of the wavefunction $\phi_0(z)$ at the top of FIG.\ref{locmin} it is apparent that modelling the ensemble approximately as a large coherent spin state, as is done in classical mean-field models, is not valid near the minimum gap; the distribution is clearly not unimodal here. It is also clearly not accurate to model the state here as a GHZ-like bimodal distribution as in Region I. A more faithful description, as we have seen, is a ground state of a quartic potential.



\section{Ground States: Regions I and III} \label{GSregionI}

Ref.\onlinecite{cirac1998quantum} proposed annealing all the way from a spin-coherent state in Region III to a `Schr\"odinger Cat' (GHZ/NOON-like) state in Region I. A similar technique was advocated in Ref.\onlinecite{zheng2002quantum} where atom-atom couplings were generated via interaction with a strong classical driving field in a thermal cavity to produce multi-atom states such as the GHZ state. At $\Gamma = 0$ the ground state is exactly a GHZ state and, as the field is ramped back up, its two delta components broaden into symmetrized ($ | \psi_{+} \rangle +  | \psi_{-} \rangle$) or anti-symmetrized ($ | \psi_{+} \rangle -  | \psi_{-} \rangle$) pairs of Gaussian lobes, FIG.\ref{pseudo}. Close to the well bottoms  $ \pm y_0$ the potential is predominantly quadratic. One must remember that this is a position-dependent mass problem and that the mass function is well-approximated by its value at $y_0$. $M(y_0) = M_{\gamma}+ O (\delta^2)$ for $\gamma < 2$, where $M_{\gamma} = 2/\gamma$. The mass increases monotonically with decreasing transverse field. To second order $V(y) =  V_0 + V''_0(y \pm y_0)^2/2$ close to turning points $\pm y_0 = \pm \sqrt{1-M^{-2}_{\gamma}} + O(\delta)$, where $V_0 = -(M_{\gamma}+M_{\gamma}^{-1})/2$ and $V''_0 = M_{\gamma}(M_{\gamma}^2 - 1)$. 

Overall we have a superposition of twin harmonic oscillators with frequency $\omega = \sqrt{M^2_\gamma -1}$, minimum gap $E_2 - E_0 = \delta \omega$ and ground state energy $E_1 \approx E_0= \delta \omega /2$, indicating the almost degeneracy between the even and odd parity eigenstates, $ | \psi_{+}  \rangle \pm | \psi_{-}  \rangle$.

This approximation at quadratic turning points has been shown robust, even outside the wells extending into much of the forbidden central barrier region \cite{garg2000tunnel}. The width $\sigma = \sqrt{\delta/(M_{\gamma} \omega)}$ of each Gaussian lobe in the $y$ variable increases monotonically with the applied field and one may write $\sigma = \Sigma/\sqrt{N}$ where coefficient $\Sigma$ is independent of $N$. The small tunneling probability through the barrier slightly lifts the energy of the anti-symmetric state $ | \psi_{+} \rangle -  | \psi_{-} \rangle$, but this gap remains exponentially small in $N$. The energy gap to the second and third excited states, also nearly degenerate, is approximately $\delta \omega$. (In fact, the Sturm-Liouville theorem guarantees that there can be \emph{no} degenerecies in a one-dimensional system, and that pairs of almost degenerate states are grouped with odd states above the even states -- see for instance Ref.\onlinecite{robnik1999wkb}.)

In the strong transverse field of Region III, the pseudo-potential $V(y)$ of eqn.\eqref{pseudopot} is dominated by its quadratic term, $V(y) \approx -1 + (1-2/\gamma)y^2/2$ and the eigenstates will be approximately those of a harmonic oscillator centered on $y=0$; therefore the effective mass remains $M_2 \approx 1$, and is no longer a function of the applied field for $\gamma >2$. The Schrodinger equation for the strong transverse field is:
\begin{align}
\left[- \frac{\delta^2}{2 M_2} \frac{d^2}{dy^2} + \frac{M_{2}}{2} \left(1 -\frac{2}{\gamma}\right) y^2 \right] \psi_n = \left( \frac{E_n}{\Gamma}+\frac{1}{M_2} \right) \psi_n
\end{align}
The unnormalized eigenstate is $\langle y | \psi_0 \rangle \propto \exp \{ - j M_2 (\sqrt{1 - 2/\gamma})y^2/2 \}$. Collecting these results,  in the thermodynamic limit ($N \gg 1$) the energy gap of $j \hat{H}$ is
\begin{subequations} \label{omegat}
\begin{align} 
\omega & = \sqrt{(\Gamma -2) (3 \Gamma - 2)} \; , & (\Gamma < \Gamma_c)&  \\ 
& = 2 \Gamma \sqrt{3 - 2/\Gamma} \; , & (\Gamma > \Gamma_c)& \; .
\end{align}
\end{subequations}

We shall see in subsequent sections how these three qualitatively very different ground states: the bimodal distribution in Region I; the broad centrally-weighted Goldilocks state in Region II;  and the Gaussian state in Region III, compare as interferometric probes. In the appendix, we examine the entanglement present during the annealing.

\section{Quantum Parameter Estimation in Presence of Noise}
It would seem that the Goldilocks state in Region II has some of the right qualitative features for metrology. To quantify the supra-classical precision in e.g. estimation of an interferometric phase $\theta$, the mean-squared error $\Delta^2 \theta$ is lower-bounded by the Cramer-Rao inequality,
\begin{equation}
\Delta^2 \theta \geq 1/(kF) \; , 
\end{equation}
where $k$ is the number of repetitions of the experiment and $F$ is the quantum Fisher information (QFI). Our objective in quantum metrology is usually to maximize this objective function $F$, which depends on both probe state $| \psi \rangle$ and the dynamics. The formalism developed in Refs. \onlinecite{knysh2014true,jiang2014quantum} presents the QFI as an exact asymptotic series in powers of $1/j$ or $1/N$. Writing $ d \psi /dy$ as $\psi'(y)$  the QFI for estimation of a phase $\theta$ associated with unitary evolution under $\hat{J}_z$ in the presence of noise has the form of a generalized `action':
\begin{equation}
\frac{F}{N^2}  =  \int_{y= -1}^{+1} \frac{\psi^2(y)}{\mu(y)} dy \:  - \:  4 \int_{y= -1}^{+1} \left( \frac{\psi'(y)}{\mu(y)} \right)^2 dy  \; , \label{action}
\end{equation}
excluding cubic and higher powers of $1/\mu(y)$, as is valid in the $N \gg 1$ asymptotic limit. The `noise function' $\mu(y) > 0$ is responsible for both effective mass and potential in the above action, and is proportional to $N$ or $N^2$ -- it depends on the type and strength of the noise present, see appendix \ref{decoh-fn}.

From  eqn. \eqref{action}, it is apparent that, for large ensembles $N \gg 1$, only those state profiles $\psi(y)$ with smoothly-varying features will be optimal. The term squared in the gradients $\psi'(y)$ has a negative sign, penalizing QFI, and therefore precision.

\section{Interferometric Performance of Goldilocks Ground State} \label{critprec}
Consider a combination of classical phase fluctuations of size $\Delta \theta = \sqrt{\kappa^0}$ and local noise  $\kappa^{(L)}$. Putting $\mu(y) = N^2 \kappa^{0} + N \kappa^{(L)} / (1-y^2)$ (more details in the appendix) into eqn.\eqref{action} means calculating terms like the second moment $\langle y^2 \rangle = \int_y y^2 \psi^2(y) (dy)$, and paying particular attention to the \emph{penalty} terms featuring squared gradients:  $\int_y \psi'(y)^2 (dy) = \delta^{-2} \langle \hat{P}^2 \rangle$, (because $\psi'(y)= i  \delta^{-1} \hat{P} \psi$). We might naively propose the `phase' state, which has $\psi_m =  1/\sqrt{N+1}$ for $m \in [-j , +j]$ as a quantum probe; it is a completely flat distribution. But it produces a large spurious gradient  $\propto \sqrt{N}$ at the boundary between $\psi_j  =  1/\sqrt{N+1}$ and $\psi_{j+1} = 0$  remembering the definition of eqn. \eqref{grad-def}. Also, the role of the probe component variance, or equivalently, the amount of `squeezing' is significant; it is not simply that more is better. Optimal probes will have  variance dictated by the strength of noise present.

For a Goldilocks state close to the critical point $\gamma_c = 2$, let us examine the dominant penalty term, $ \delta^{-2} \langle \hat{P}^2 \rangle$.  For eigenstates of $H$ it is easy to show $\langle [\hat{H},\hat{P} y]\rangle = 0$ and calculating this commutator\footnote{Remembering $[y,\hat{P}] = i  \delta$ and the inner derivatives: $[Q(y),\hat{P}] = i \delta \frac{d Q(y)}{d y}$ and $[R(\hat{P}),y] = - i \frac{d R(\hat{P})}{d \hat{P}}$} provides an expression for $\langle \hat{P}^2 \rangle$. Expanding to $y^4$ near $y=0$ and converting to the scale-free variable $z$ gives:
\begin{align}
 \frac{\langle \hat{P}^2 \rangle}{\delta^2}&= 4 \left(1- \frac{2}{\gamma}\right) g^{2/3}  \langle z^2 \rangle + 2 g^{1/3} \langle z^4 \rangle  \nonumber \\
 & = g^{1/3 } \left( a \langle z^2 \rangle + 2 \langle z^4 \rangle \right)
\end{align}
But we also have, from the Schr\"odinger equation, that $\langle \hat{P}^2 \rangle/\delta^2= g^{1/3}[\epsilon_n(a) - a \langle z^2 \rangle - \langle z^4 \rangle  ]$ so we can eliminate $\langle z^4 \rangle \mapsto (\epsilon_n(a) - 2 \langle z^2 \rangle)/3$. Identifying $\epsilon'(a) = d \epsilon(a)/da = \langle z^2 \rangle$ using the Hellmann-Feynman theorem for parameter $a$ gives the precision penalty factor (PPF):
\begin{equation}\label{penalty}
\frac{\langle \hat{P}^2 \rangle}{\delta^2}=\frac{g^{1/3 }}{3} \bigg( 2 \epsilon_0(a) - a \epsilon_{0}'(a)\bigg) \; .
\end{equation}

An exact numerical search reveals $a \mapsto a_F \approx -2.5536$. This parameter value $a_F$ minimizes the PPF above at $1.4239 g^{1/3}/3$; this is analogous to how the location $a_0$ of the minimum gap was found earlier, see the right side plot of FIG. \ref{wavy}. Scaled back to the annealing variable $\Gamma$, optimal precision occurs at:
\begin{equation}\label{gf}
\Gamma_F = \Gamma_c \left(1-\frac{a_F}{12 g^{1/3}}\right)^{-1} = \Gamma_c \left( 1-\frac{a_F}{3} \left(\frac{1}{2 M_2 N}\right)^{2/3} \right)^{-1}
\end{equation}
Note that this optimum point on the annealing schedule is not a function of the decoherence strength or type -- it depends only on the number of qubits (to leading order).

The expansion of minimum error is:
\begin{align}\label{region2error}
\left[ \Delta^2 \theta \right]_{\text{II}} \geq & \; \frac{1}{F(\Gamma_F)} \;  = \kappa^{0} + \frac{\: \kappa^{(L)}}{N} \nonumber \\
 + & \; \frac{2}{3 N^{4/3}} \left( \frac{M_2}{2}\right)^{\frac{1}{3}}\bigg( 2 \epsilon_0(a_F) - a_F  \epsilon'_0(a_F) \bigg)
\end{align}
ignoring terms $O\left(1/N^{5/3} \right)$ and smaller. The leading two terms are independent phase errors from collective phase noise and local noise that has shot-noise scaling $\propto 1/N$. Together, they represent the \emph{ultimate} upper bound to precision. The next significant term, in $1/N^{4/3}$, would provide the leading $N$ dependence in the absence of local noise. It also dictates how fast the upper bound $ \kappa^{0} + \frac{\: \kappa^{(L)}}{N}$ may be approached (the more negative the power of $N$ in the third term above, the faster the convergence). Note that both $\kappa^{0}$ and $\kappa^{(L)}$ are absent from this term $\propto N^{-4/3}$, its contribution to phase error comes from $\langle \hat{P}^2 \rangle$ alone, i.e. the probe shape. The generic $1/N$ scaling of precision for most quantum channels was first derived in Ref.(\onlinecite{fujiwara_fibre_2008}) and later, optimal probe shapes were found for lossy interferometry \cite{knysh_scaling_2011}.

\begin{figure}
	\includegraphics[width=3.4in]{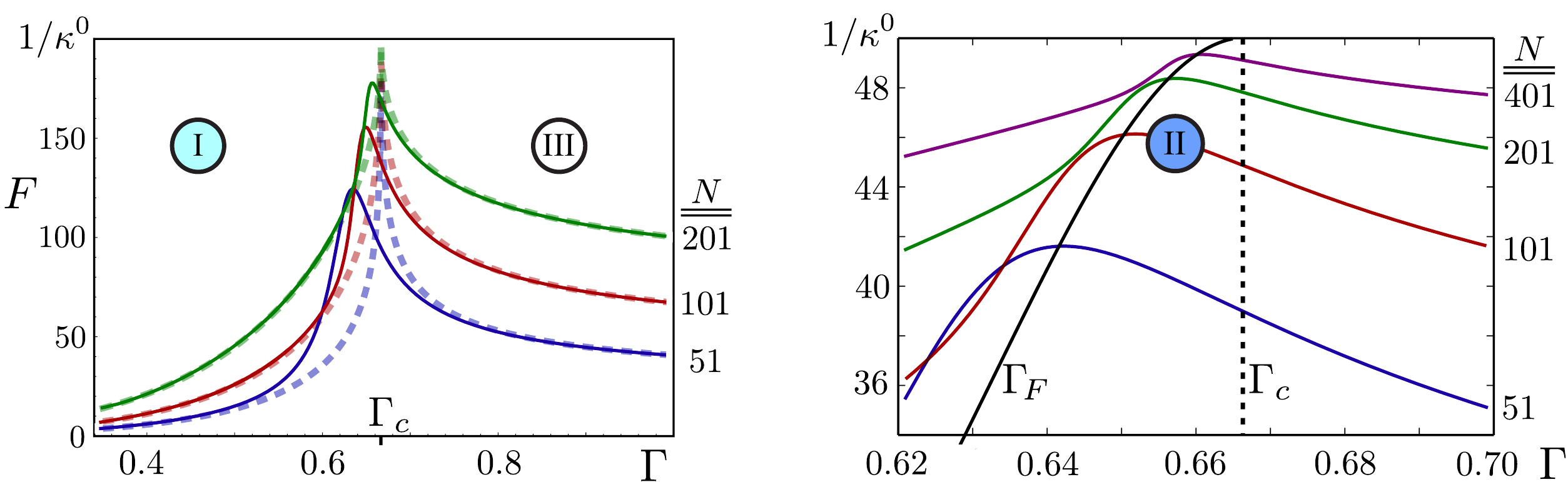}
	\caption{\label{rgb}
		\textbf{Numerics and Analytics:} Dashed curves show precision $F(\Gamma)$ for analytic ground states of Regions I and III in the thermodynamic limit; states discussed in section \ref{GSregionI} and precision given in eqn.\eqref{Finf}. Here $N = 51,101,201$ qubit ensembles (blue, red and green curves) are subjected to collective dephasing $\kappa^{0}= 1/200$. Exact numerical results are unbroken curves of the same colour. First, ground states of the original spin problem are found by direct matrix diagonalization. Second, the QFI is calculated by additional diagonalization of the mixed state to which the ground state evolves under decoherence, see eqn. \eqref{qfi-diag} in the appendix.  The discrepancy between dashed/unbroken curves in the left figure is due to the absence of Goldilocks Region II in the thermodynamic limit. The right plot zooms in on the critical region for larger dephasing $\kappa^0 = 1/50$ and $N \in \{51,101, 201, 401 \}$. Coloured curves are still numerical results of QFI. The unbroken black curve gives the analytic locus of the predicted QFI maxima $\{\Gamma_F, F(\Gamma_F)\}$ for all $N$, using the asymptotic results of eqns. \eqref{gf} and \eqref{region2error}. The formula for $F(\Gamma_F)$ is only valid when the condition $\kappa^0 N^2 \gg 1$ is met, as it is in this case.}
\end{figure}

\section{Interferometric Performance of Ground State in Strong and Weak Fields}
To contrast, for increasing transverse field $\Gamma > 2/3$ the ground-state is an approximately Gaussian-distributed wavefunction $\psi(y) \propto \exp \{ - j M_2 (\sqrt{1 - 2/\gamma})y^2/2 \}$, becoming eventually a spin-coherent state aligned with the field in the spatial $x$-direction, a separable state. For $1/F$ one recovers an exact expression for the lower bound: 
\begin{equation}
\left[ \Delta^2 \theta \right]_{\text{III}} \geq \kappa^{0} + \left(M_2 \sqrt{3-2/\Gamma}+\kappa^{(L)}\right)/N \; .
\end{equation} 
It would appear that the ultimate upper bound $\kappa^{0}+ \kappa^{(L)}/N$ can be approached in the limit $\Gamma \sim 2/3$; the associated Gaussian wavepacket would, however, have infinite variance. But for $\kappa^{(L)} < 1$ and small $\kappa^{0} <  \kappa^{(L)}/N$, there exists a good argument for preparing the probe state as close to $\Gamma_c$ as possible. The blow-up in QFI predicted above for Region III at criticality, combined with the more accurate results derived previously for Region II, together promote the vicinity of $\Gamma_c$ as optimal for probe preparation. But what about Region I?

\begin{figure}[!]
	\includegraphics[width=3.2in]{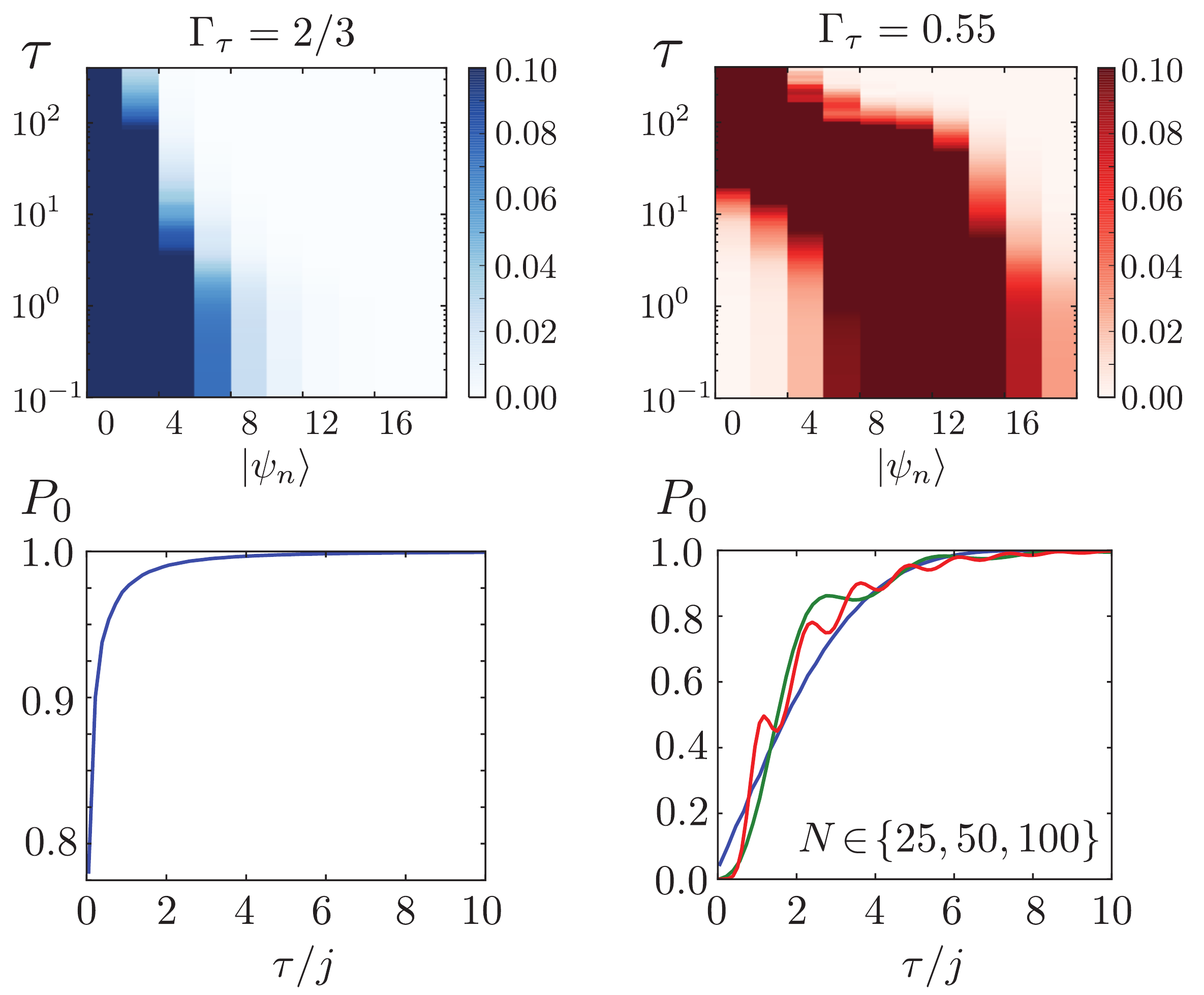}
	\caption{\label{anneal-time}
		\textbf{Annealing:} Upper plots show for $N=100$ the overlap $|\langle \psi|\psi_n \rangle|$of the annealed state with the $20$ lowest energy states $| \psi_n \rangle$  with $n \in \{ 0,19\}$ subject to a linear annealing schedule of total time $\tau$, so $\Gamma(t) = 1- (1-\Gamma_\tau) t / \tau$. Odd-numbered energy eigenstates are excluded because their overlap with the annealed state is zero at all times. They have odd parity and the annealing Hamiltonian respects the even parity of the initial state at $t=0$. On the left, the annealing schedule is terminated at the critical point $\Gamma_\tau = \Gamma_c$, and on the right, it is terminated some way into Region I after traversing the minimum gap. When the annealing is halted in the critical region, the instantaneous ground state is still dominant across many orders of annealing time. The lower two plots show the probability $P_0$ to be in the target ground state, again for an annealing cycle halted at $\Gamma = 2/3$ and $0.55$.  The horizontal axis is the total annealing time in the linear schedule,  in units scaled by the number of qubits: $\tau/N$. The blue, green and red curves are for systems of $25, 50$ and $100$ qubits, respectively. (In the bottom left plot, the three curves are identical.) Annealing success for a simple linear schedule apparently scales linearly with ensemble size, as discussed in section \ref{annealtime}.}
\end{figure}

In Region I, transitions between the ground state $ | \psi_{+} \rangle +  | \psi_{-} \rangle$ and first excited state $ | \psi_{+} \rangle -  | \psi_{-} \rangle$ are prohibited, due to their opposite parity and because evolution via the time-dependent Schr\"odinger equation is parity-conserving. Unfortunately, any noise or decoherence is \emph{unlikely} to respect parity, so even at very low temperatures, thermalization occurs to an equal mixture of the two near-degenerate states. Such an effectively 2-level (qubit) maximally-mixed state will be symmetric under unitary evolution by the interferometric $\hat{J}_z$ operator, and useless in estimating the associated phase parameter\cite{javanainen2014ground}.

Let us imagine that the symmetrized superposition of gaussians could be prepared adiabatically. For hybrid noise,  the QFI `action'  integral of eqn.\eqref{action} produces:
\begin{align}
\left[ \Delta^2 \theta \right]_{\text{I}} \geq \kappa^{0} +\frac{1}{N} \left\{\kappa^{(L)} (1+ y_0^2) + \frac{1}{\Sigma^2}\right\} + O\left( \frac{1}{N^2}\right) \nonumber \\ =  \kappa^{0} +\frac{1}{N} \left\{\kappa^{(L)} \left(2 - \frac{1}{M_\gamma^{2}}\right) + M_\gamma \sqrt{M_\gamma^2-1}  \right\} + O\left( \frac{1}{N^2}\right)
\end{align}
approximately valid as long as the wells retain a parabolic shape across the width of the ground-state lobes. Notice that, for this noisy scenario, error increases with the width of the central barrier separating the two wells $2 y_0$ (shown in FIG.\ref{pseudo}) and there is a additional precision penalty for narrower Gaussian lobes of width $\sigma = \Sigma/\sqrt{N}$; the opposite behaviour is seen in a noiseless environment, where QFI increases quadratically with the ratio $y_0/\sigma$. Apparently, wider lobes that are closer together (less `cat-like') improve robustness to noise.  

In the limit $\Gamma \sim 0$, the assumption of smoothly varying ground state amplitudes is no longer valid-- the state  is in reality a NOON or GHZ state, whose interferometric performance in the presence of noise has been shown elsewhere to scale exponentially badly in ensemble size $N$. For collective dephasing  QFI is $N^2 \exp \{- \kappa^0 N^2 \}$ and in dissipative systems of transmission $\eta<1 $ it is $\text{N}^2 \eta^N$. (Refs.\onlinecite{bardhan2013effects,dorner2009optimal}).

The unbroken curves in FIG.\ref{rgb} show the performance $F(\Gamma)$ calculated numerically for the original spin system in the presence of interferometric phase noise. These curves asymptote for $N \gg 1$ to give the analytical result in the thermodynamic limit :
\begin{equation}\label{Finf}
F_{\infty}(\omega)=  \left( \kappa^{0} + \frac{\kappa^{(L)}}{N} + \frac{M\omega}{N} \right)^{-1}  \; , 
\end{equation}
where  $\omega$ is given in eqn.\eqref{omegat} and  $M \mapsto M_{\gamma} = 2/\gamma$ for $\gamma < \gamma_c$ and $M \mapsto M_2\approx 1$ for $\gamma > \gamma_c$. The asymptotes ignore the critical region entirely; recall that it vanishes in $\Delta \Gamma$ near $\Gamma_c$ at a rate $\propto 1/N^{2/3}$.

Comparing $1/F_{\infty}(\omega)$ with eqn.\eqref{region2error} we see explicitly that the ultimate precision limit $\kappa^{0} +  \kappa^{(L)}/N$ is \emph{only} asymptotically saturable in Region II; in the other regions there is an additional contribution to mean squared error of order $1/N$ proportional to the gap $\omega$. This is a central result of this paper.

\section{Annealing Time Complexity} \label{annealtime}

Annealing time can depend on the requirement of adiabaticity -- whether the system needs to be in the instantaneous ground state at all times. If this can be relaxed, the annealing time can be reduced. Roughly speaking, the annealing schedule must progress slowly when the gap between the ground and first excited state is small. The exponential scaling in $N$ of the time complexity of certain quantum algorithms can be traced to an exponentially small minimum gap. In the current context, the gaps for $j \hat{H}$ in Regions I and III are fixed and independent of $j$ or $N$ (derived from the ground state approximation calculated in section \ref{GSregionI}). That leaves only Region II. Choosing the final annealing parameter as $\Gamma_\tau$ and using a prescription from Ref. \onlinecite{van2001powerful}, an optimal annealing time $T$  can be calculated as
\begin{align}
T &\approx \int_{\Gamma_\tau}^{1} \;\left|\left| \frac{d (j \hat{H})}{d \Gamma} \right|\right|_2 \; \frac{d \Gamma}{\omega^2(\Gamma)}  \; \; ,
\end{align}
where $|| M ||_2$ is the $2$-norm of a matrix $M$. For the current Hamiltonian $
\left|\left|  \frac{d (j \hat{H})}{d \Gamma} \right|\right|_2 = \left|\left|  \hat{J}_z^2 / j - \hat{J}_x  \right|\right|_2 \; \sim \; 5j/4$ for $j \gg 1$. 
This matrix norm factor is linear in $N$ and independent of $\Gamma$ throughout all regions. As presented in eqns. \eqref{omegat}, irrespective of where the annealing is halted the contributions from $1/ \omega^2$ in both Regions I and III approaches a constant, leaving only the calculation for Region II. There,  $\Gamma_\tau \subset \Gamma_c \pm \Delta \Gamma_G$, and the gap $\omega_{\text{II}} = \Delta E_{20} \propto j^{2/3}/\Gamma$ from eqn. \eqref{gap}. Then we have an annealing time $T _{\text{II}}\approx$
\begin{align}
 \frac{5 j}{4}  \int_{\Gamma_c - \Delta \Gamma_G}^{\Gamma_c + \Delta \Gamma_G}  \frac{d \Gamma}{\omega_{\text{II}}^2(\Gamma)} \propto j^{\frac{5}{3}} \int_{\Gamma_c - \Delta \Gamma_G}^{\Gamma_c + \Delta \Gamma_G}  \frac{d \Gamma}{\Gamma^2} \sim j^{\frac{5}{3}} \frac{\Delta \Gamma_G}{\Gamma_c^2} \propto j
\end{align}
where, in the last step, we have recalled that the Goldilocks zone scales $\Delta \Gamma_G \sim j^{-2/3}$.

 Overall time complexity is  $T \sim O(j)$ in all three annealing regions. This estimate could be considered pessimistic as it applies only to adiabatic passage. For a linear annealing schedule, numerical results confirm that, irrespective of where the annealing is halted (near criticality or all the way to the weak field limit), the annealing time for any desired fidelity to the target ground state is linear in $N$, see FIG.\ref{anneal-time}. Whether terminating at a GHZ-like state or Goldilocks probe, the total time differs only by a fixed factor independent of $N$. Since the annealing time scales favorably with the size of the ensemble, then decoherence may be less significant during probe preparation. 


\section{Convergence on Asymptotics}
\begin{figure}[t!]
	\includegraphics[width=2.8in]{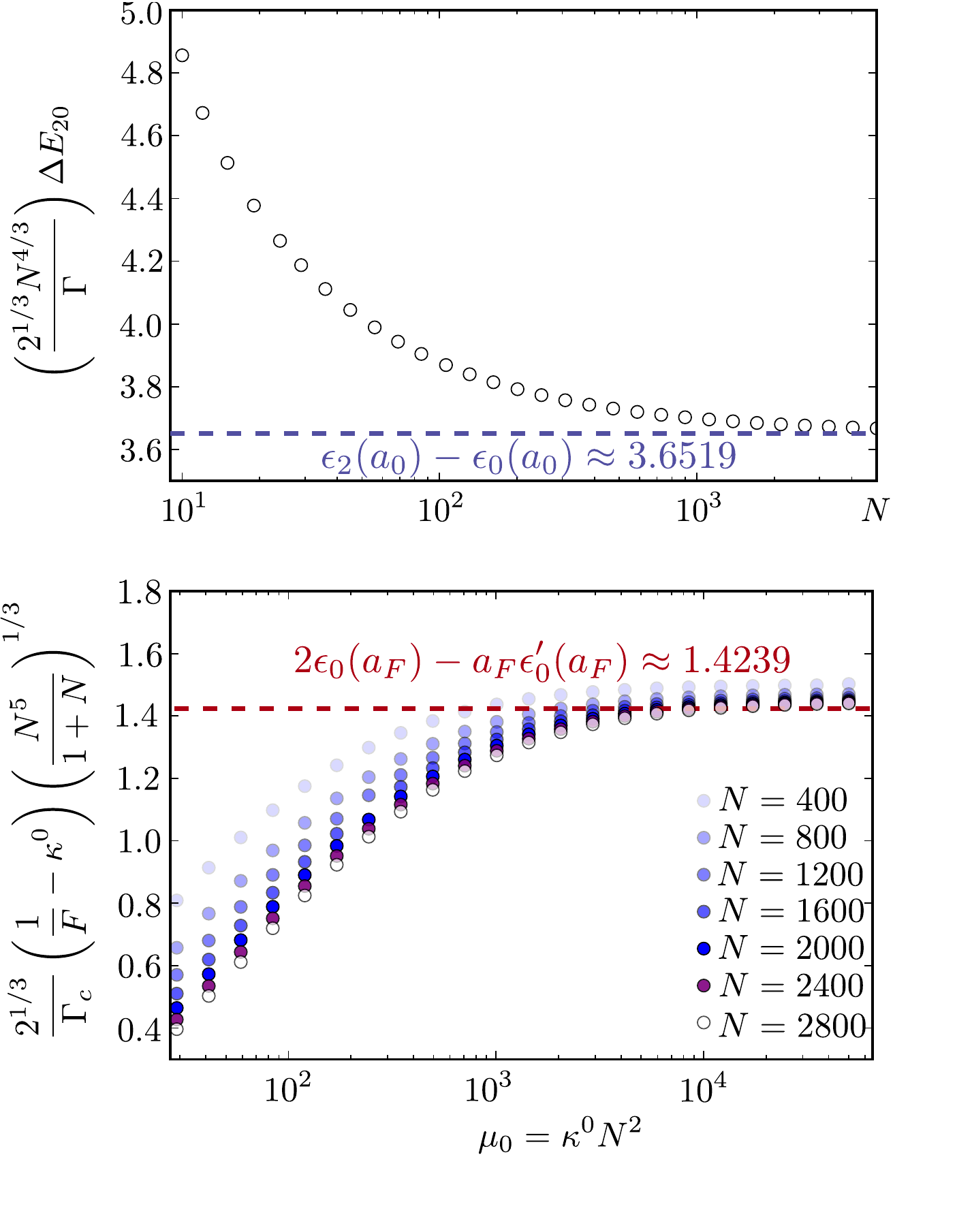}
	\caption{\label{ppf}
		\textbf{Convergence to Asymptotics:} Vertical axes plot pure numerical factors expected to converge on $\epsilon_2(a_0) - \epsilon_0(a_0) $ (upper plot) and $2 \epsilon_0(a_F) -a_F \epsilon'_0(a_F)$ (lower plot), a re-arrangement of eqn. \eqref{gap} and eqn.\eqref{region2error}. The asymptotic predictions are $3.6519$ and $1.4239$, respectively for the gap and penalty factor (blue and red dashed lines). Since these derive from the scale-free picture, they are independent of both $\Gamma$ and $N$.  }
\end{figure}

In the scale-free setting of Region II, the pure numerical values of minimum gap and precision penalty factor are predicted to be $\epsilon_2(a_0) - \epsilon_0(a_0) $ and $2 \epsilon_0(a_F) -a_F \epsilon'_0(a_F)$, respectively. This latter minimum corresponds to maximum QFI. If instead the maximum QFI is found by brute force numerical diagonalization and eqn.\eqref{qfi-diag} for different ensemble sizes $N$ and collective phase noise amplitude $\kappa^0$, the true penalty factor can be determined including all finite-size corrections. The ground-state gap in the original spin dynamics $\Delta E_{20}$ is also easily converted into that of the scale-free setting. The convergence to the predicted asymptotic values is seen in FIG.\ref{ppf}. From these graphs it is clear that convergence is fairly slow, but by $N \approx 10^3$ and  $\mu_0 > 10^4$ the numerical values do approach these asymptotic bounds. This observation provides crucial evidence in validating the sequence of approximations that have been made; mapping from a spin system to a 1-D particle in a potential, restriction to a quartic potential in the critical region, and approximation of the QFI by its two leading terms in the exact asymptotic series. The reason for the slow convergence could be attributed to the first  term excluded from the QFI series being only slightly smaller than the last included term, $N^{-5/3}$ vs $N^{-4/3}$. The value of $N$ has to become quite large before $N^{-4/3}$ can dominate.

\section{Conclusions and Outlook}

We have examined a network of uniformly-coupled spins in a transverse field as an interferometric probe for use in noisy conditions. Mapping the ensemble onto a variable-mass particle in a potential allowed quantitative understanding of the dynamics in the critical region, i.e. we were able to characterize correctly the dominant properties of the continuous phase transition. In terms of  annealing parameter $\Gamma$, we discovered the ordering and distances between the critical point in the thermodynamic limit $\Gamma_c$, the minimum gap $\Gamma_0$ and the point of maximum precision $\Gamma_F$ using an exact numerical approach. These latter landmarks are the relevant ones for annealing and metrology, and are sufficiently far from $\Gamma_c$ that conventional perturbative techniques would fail. Utilizing asymptotic formulae for QFI, we predicted that in a noisy environment, best precision is offered \emph{only} by ground states prepared near the critical point. We saw in eqn.\eqref{region2error} that such states asymptotically \emph{saturate} the ultimate precision bounds for interferometers subjected to typical noisy environments. We confirmed the accuracy of both the asymptotic QFI expansion and the legitimacy of the continuous particle mapping by brute force matrix diagonalization methods for the original spin system, for $N \in [30,3000]$. (In the asymptotic approach, only the combined parameter $\mu$ is required to be large -- the product of noise strength and ensemble size.)  We determined that adiabatic probe preparation has a time-complexity scaling linearly with ensemble size. (In the appendix, the precision of the $N$-qubit Goldilocks probe is compared numerically with probes prepared by a sudden quench of the transverse field.)

In Ref. \onlinecite{knysh2014true} the calculus of variations dictated that asymptotically the best-performing interferometric state was \emph{always} the ground state of a 1D particle in a special pseudo-potential, created between two repulsive Coulomb sources and identical to the noise function $\mu(y)$. In some sense,  we have tried to engineer non-linear dynamics that best mimic that optimal potential. Although a quartic potential does not much resemble the optimal one, any ground state of width in the variable $y$ that narrows with increasing $N$ will not `explore' the structure of the potential far from $y=0$; such a probe can have some of the desired properties in the large $N$ limit. 

Decoherence during probe preparation must be strongly suppressed, e.g. fluctuations in the transverse field $\gamma$ are amplified in the $a$ variable\footnote{Assuming $\Gamma$ is a Gaussian-distributed random variable of width $\sigma$, then $a$ is also a random variable, but with a non-gaussian distribution $q(a) = p(a) \left| \frac{d \Gamma}{da} \right|$ that looks increasingly gaussian for $N \gg 1$ and fluctuations near the mean. The width in $a$ is now approximately $\sigma g^{1/3}/18$. Remember the interesting range of $a \in [-4,0]$ so the transverse field noise must be suppressed by a factor $N^{2/3}$ if the Goldilocks zone is to be located at all.}  by $N^{2/3}$ -- the strength of these  fluctuations places an upper limit on the size of the spin ensemble that can be prepared in the critical region. The full effects of decoherence during the annealing schedule we leave to a future publication.

The results of this paper promote an alternative perspective on the developing technology that is the quantum annealing machine, e.g. the pioneering work of Ref.\onlinecite{johnson2011quantum}. Typically, the goal of such devices is to prepare a ground state that represents the optimal solution to a combinatorial problem encoded directly in the couplings of an Ising Hamiltonian. Here, it has been proposed that such customizable dynamics might instead be used to prepare some exotic yet useful quantum state of many qubits. Perhaps the application is metrology as discussed; other possibilities include quantum communication, or generation of different types of entanglement\cite{lanting2014entanglement} for distributed quantum information. Presented in the context of ion traps and optical lattices, the authors of Ref. \onlinecite{m2000risq} had already recognized the potential of a Dicke-Ising model of quantum computing for simulation of quantum systems, and as a resource for generating squeezing and entanglement. More recently, the creation of tunable Ising systems optically has been proposed in QED cavities\cite{gopalakrishnan2011frustration} but not yet considered for metrology applications.

The inverted challenge in terms of global optimization is to `reverse engineer' the Ising couplings to prepare a particular known ground state of interest. Now, the search objective is the associated Hamiltonian couplings and topology. When preparation time is a significant resource, one may have to offset state fidelity against shorter annealing times, if the landscape necessitates annealing through gap regions, or if adiabaticity is not a strict requirement. Adding external control fields might avoid proximity to the smallest gaps, and allow adiabatic short cuts\cite{torrontegui2013shortcuts} such as transitionless driving\cite{berry2009transitionless}.

\appendix

\section{Variable-Mass Schrodinger Ground State compared numerically with that of Spin System}
\begin{figure}
	\includegraphics[width=1.8in]{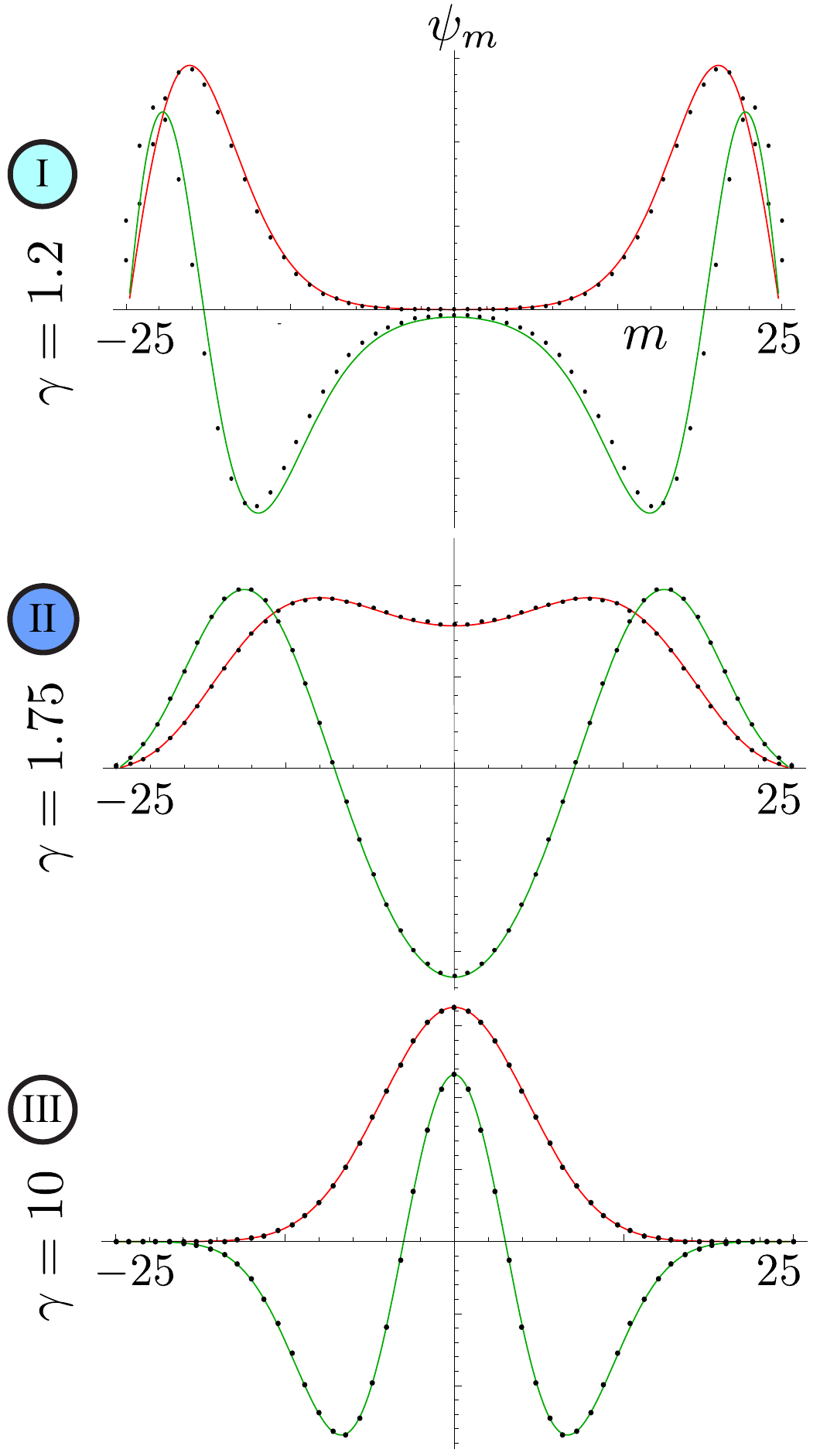}
	\caption{\label{greenred}
	For the three annealing regions, (I: weak field), (II: critical field), (III: strong field), discrete ground state amplitudes and second excited state amplitudes for a system of  $50$ qubits $(N = 2j)$ are indicated by black dots. The equivalent continuous wavefunction eigenstates for the variable-mass particle are shown as unbroken green (ground state) and red (second excited state) lines.}
\end{figure}

Solutions to eqn.\eqref{schrod1d} are shown in FIG.\ref{greenred}  for the ground (red) and second excited state (green)  for the variable-mass particle in the one-dimensional potential $V(y)$ and different values of the annealing ratio $0<\gamma< \infty$ (remembering the critical point is at $\gamma_c =2$). Compared the discrete ground state amplitudes $\psi_m$ for the original quadratic spin Hamiltonian, good agreement is obtained.  Fidelity to the original Hamiltonian eigenstates improves with larger ensembles and larger values of $\gamma$. For $\gamma \gtrsim 0$ the spin eigenstates contain discrete delta-like components (GHZ state) and the continuous approximation is no longer valid. Also, the continuous variable solution depends on the boundary conditions; we have chosen $\psi(y) = 0$ at $y = \pm 1$ but the discrete amplitude set $\{ \psi_m\}$ can be non-zero at $m = \pm j$ for finite $j$. An additional difference between the models is the discrete number of eigenstates for the spin system -- in contrast, the particle model has no upper bound to the number of excited states.

\section{Global Entanglement}
Partial entanglement is necessary for probes to offer supra-classical precision in noisy interferometry. It will be a useful exercise,  therefore, to quantify the entanglement present during the annealing process, in particular through the phase transition. To characterize entanglement a useful measure is the \emph{global geometric entanglement} \cite{chen2014comparison} (gge) which we can define for a pure entangled state $|\psi \rangle$ as:
\begin{equation}
\mathcal{G}[|  \psi \rangle] = \text{min} \left\{ - \log_2 |\langle \chi|  \psi \rangle|^2 \right\} \; ,  \; \; \; \forall \; | \chi \rangle \in S
\end{equation}
i.e. where $| \chi \rangle$ belongs to the set of separable states $S$ and the minimization is performed over all $S$. The function $\mathcal{G}$ is the negative logarithm of the fidelity of the entangled state to the nearest separable state. The nearest separable state will in fact be a pure (product) state since the ground state is pure. The gge is sensitive to bipartite and multi-partite entanglement, although it does not differentiate between different entanglement depths.

For the current dynamics the spin ensemble lives in the maximum spin sector ($j_{\text{max}} = N/2$), and therefore the ground state is fully permutation-symmetric. It is simple to argue that the nearest separable state also shares this permutation symmetry. If part of that state lived in a different $j$ sector, an orthogonal subspace, it would only \emph{reduce} the overlap $|\langle \chi | \psi \rangle|$. The only fully-symmetric pure separable states are in fact the spin-coherent states $|\chi \rangle \mapsto | \alpha, \beta \rangle$. Finding the gge can be a difficult optimization in general, but for symmetric states it means finding the optimal $\alpha,\beta$ angle pair -- the polar and azimuthal angles of the spin vector giving maximum overlap with $| \psi \rangle$. The spin-coherent state\cite{barnett2002methods} has components:
\begin{equation}
\langle m |  \alpha, \beta \rangle = \left(\cos\frac{\alpha}{2} \right)^{2 j} \sqrt{\binom{2 j}{j+m}} \left(\ee^{- i \beta} \tan \frac{\alpha}{2}  \right)^{j+m}
\end{equation}
The ground state of the quadratic spin system has real coefficients, as does the closest spin coherent state: $\beta = 0$. Also the probability distribution $|\langle m |  \alpha, \beta \rangle|^2$ is binomial, with mean $\langle \hat{J}_z \rangle = j \cos \alpha$ and variance $(\Delta \hat{J}_z)^2 = (j/2) \sin^2 \alpha$. Approximating the binomial distribution as Gaussian in the $j \gg 1$ limit and converting to $y$ produces a mean $y_\alpha = \cos \alpha$ and standard deviation $\sigma_\alpha = j^{-1/2} |\sin \alpha|$. Obviously $\sigma_\alpha \leq j^{-1/2}$ is the upper bound on wavefunction `width' for a separable state.

In Region III the ground state is centered on $y=0$ and the nearest spin-coherent state will also be centered on the origin, being as wide as possible, i.e. $\alpha = \pi/2$. The squared overlap of two Gaussian wavefunctions with the same mean but different variances is $2 \sigma_a \sigma_b/(\sigma^2_a + \sigma_b^2)$. The gge for $\Gamma > \Gamma_c$ is then:
\begin{equation}
\mathcal{G} = \log_2 \left[  (3 -2/\Gamma)^{1/4} +  (3 -2/\Gamma)^{-1/4}  \right] - 1
\end{equation}

As we have seen, the ground-state passing into Region I during an annealing cycle bifurcates into two approximately Gaussian lobes. The nearest spin-coherent state will choose one of those lobes and attempt to match both its mean and variance (to achieve maximum fidelity). Interestingly, an almost exact matching for both quantities is possible although the spin coherent state is a function of a single parameter $\alpha$. The bi-modal lobes of the ground state have means $\pm y_0 = \pm \sqrt{1-1/M^2_{\gamma}}$ and standard deviation $\sqrt{1/y_0 - y_0}$. Fidelity to the spin-coherent state, with  $y_\alpha$ and $\sigma_\alpha$ given above, can be close to unity only if:
\begin{equation}
\alpha_{\text{opt}} =\arctan \left[ \frac{\sqrt{M_\gamma}}{(M^2_{\gamma}-1)^{3/4}} \right] \; , \; \; 0<\gamma < 2
\end{equation}
remembering that $M_\gamma = 2/\gamma = 2 \Gamma/(1-\Gamma)$.  This gives an asymptotic result for the entanglement  in terms of the variable mass:
\begin{equation}
\mathcal{G} = \log_2 \left[ \left(1-1/M^2_\gamma \right)^{1/4}  + \left(1-1/M^2_\gamma \right)^{-1/4}  \right]
\end{equation}
valid in the parameter range, $\Gamma < \Gamma_c$.
Only half the probability (with some exponentially small correction) is concentrated in a single lobe, so the overall fidelity to the ground state will quickly converge on $1/\sqrt{2}$ in Region I. Then $\mathcal{G} \sim 1$, which is the known gge for a GHZ state, a state with \emph{only} $N$-partite entanglement. This analysis also indicates how well-conceived is the model of a `cat' state with two superposed spin-coherent states, for this spin Hamiltonian and $\Gamma < \Gamma_c $, as presented in Ref. \onlinecite{cirac1998quantum}.

\begin{figure}[h!]
	\includegraphics[width=2.2in]{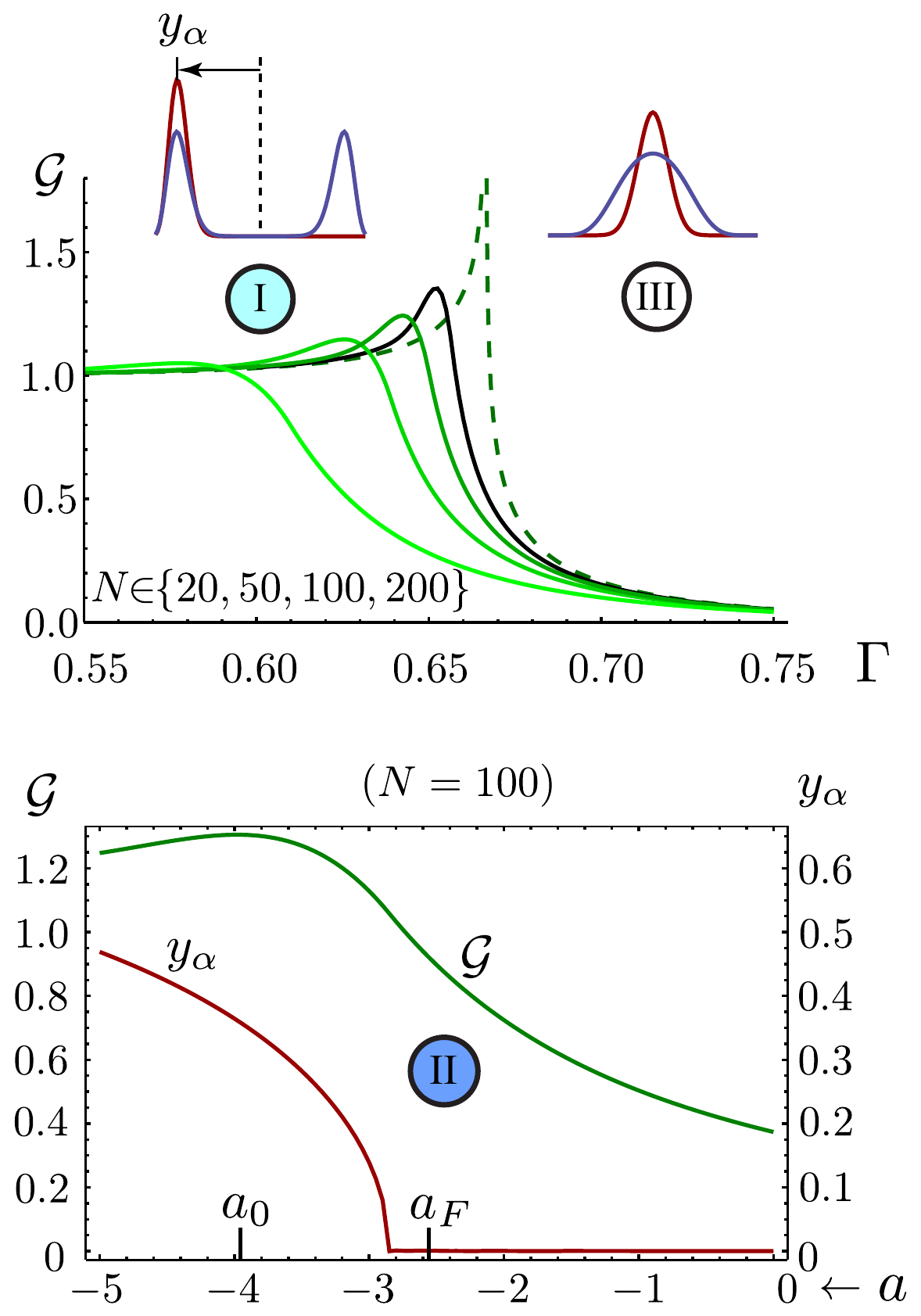}
	\caption{\label{global-ent}
		\textbf{Global Entanglement:} Upper plot shows the analytic bounds on global geometric entanglement $\mathcal{G}$ in the thermodynamic limit (dashed green curve) as a function of $\Gamma$. The unbroken green curves depict the entanglement for finite sized ensembles up to $N=200$ (black curve). The superimposed red and blue distributions show the amplitudes of the nearest spin-coherent state (red) to the ground state of the annealed system (blue) in the original spin problem. This spin-coherent state is also the nearest (or highest fidelity) separable state. Finding this state is crucial, due to $\mathcal{G}$ being the negative logarithm of the square of the overlap between these two states. When the ground state centered on $y=0$ bifurcates into two lobes as it passes from Region III to Region I the nearest spin-coherent state (having an approximately Gaussian distribution) can only track one of the lobes; its mean $y_\alpha = \cos \alpha$ starts to move with decreasing $\Gamma < 2/3$. As $\Gamma$ approaches $0$ the entanglement asymptotes to $\mathcal{G} = 1$, the entanglement of a GHZ state. The lower graph plots the entanglement more closely in Region II for a system of $N=100$ qubits in the scale-free setting. Maximum entanglement is seen to occur in the vicinity of the minimum gap, but this maximum will be closer to $a =0$ or $\Gamma = \Gamma_c$ for larger $N$. As $N$ increases the maximum width of a spin-coherent state scales as $N^{-1/6}$ in the $z$ variable. Thus maximum entanglement grows slowly as $\frac{1}{6}\log N$. The red curve plots the locus of the mean of the nearest spin-coherent state. Apparently, as $a$ or $\Gamma$ decreases the ground state has already begun to bifurcate before maximum entanglement is reached.}
\end{figure}

In the scale-free setting of FIG.\ref{global-ent} it is seen that for $N=100$ qubits the maximum global entanglement is at $a_G \approx -3.85414$. For larger $N$ or $j$ this maximum will therefore occur in the region $a_G < a < 0$ because of the properties of the nearest spin-coherent state. In the $y$ variable this state has width $1/\sqrt{j}$ but in the scale-free variable $z = y g^{1/6} \propto y j^{1/3}$, the maximum width becomes $j^{-1/6}$. The squared overlap of the widest spin-coherent state with the ground state in Region II (whose variance is just a pure number in the $z$-variable near critical annealing) is going to scale asymptotically as $\propto 1/j^{1/6}$ or $1/N^{1/6}$. The gge in the limit $N \gg 1$ therefore approaches:
\begin{equation}
\mathcal{G}_{\infty} \sim \frac{1}{6}\log_2 N
\end{equation}
which confirms the central result of Ref.\onlinecite{orus2008equivalence} for the (isotropic) Lipkin-Meshkov-Glick model. Note that the entanglement per copy vanishes in the thermodynamic limit, as $(\log N) /N \sim 0$. We should not be too shocked by this scaling law as it has been shown that although in general $\mathcal{G} < N-1$, the maximum entanglement for \emph{symmetric} states is $\mathcal{G}_{\infty} \sim \log_2 (N+1)$.

\section{Calculating Quantum Fisher Information}
Consider a phase parameter $\theta$ encoded by a spin Hamiltonian, e.g. $\hat{J}_z$ aligned with the spatial $z$ direction, acting on a noisy mixed quantum state $\rho$, of $N$ qubits. Assume the noise process commutes with the phase rotation, as is the case e.g. dissipation and for collective dephasing. The mixed state is transformed by $\exp \{- i \theta \hat{J}_z\}$ and for such finite-dimensional systems the calculation of QFI typically involves diagonalization of the density matrix \cite{paris2009quantum,toth2014quantum}. For $\rho(\theta) = \sum_i \lambda_i |\psi_i \rangle \langle \psi_i |$ then defining the QFI as $F(\theta)$, it is:
\begin{equation}\label{qfi-diag}
2 \nsum_{i,j}\frac{\left| \langle| \psi_i|\rho'(\theta)| \psi_j \rangle \right|^2}{\lambda_i + \lambda_j} \mapsto 2 \nsum_{i,j}\frac{(\lambda_i - \lambda_j)^2 }{\lambda_i + \lambda_j} \left| \langle \psi_i| \hat{J}_z| \psi_j \rangle \right|^2
\end{equation}
where $\rho'(\theta) = d \rho / d \theta$. The computation becomes increasingly arduous for $N \gg 1$ without introducing any insight into the result. Recently, a different formulation was proposed, useful in the large $N$ case, where $F(\theta)$ may be expanded as an exact asymptotic series  \cite{knysh2013estimation,jiang2014quantum}:

\begin{equation}\label{qfiseries}
F = \left\langle \left[\hat{J}_z,2 \tanh \left( \frac{1}{2} \bigg[- \log(\rho), \bullet\bigg] \right)\hat{J}_z\right]\right\rangle \: .
\end{equation}

Here square brackets denote an operator commutator, $[A,B] = A B - B A$, and angular brackets indicate an expectation value taken with the density matrix, $\langle \hat{A} \rangle = \Tr (\rho \hat{A})$. The adjoint endomorphism $[A,\bullet] = \ad_A$ acts as a superoperator. Thus $[A,\bullet]\hat{J}_z=[A,\hat{J}_z]$, and $[A,\bullet]^3 \hat{J}_z=[A[A,[A,\hat{J}_z]]]$ are terms in the power series expansion of the hyperbolic tangent; $\tanh x = x - x^3/3 + \dots$ with $x = [A,\bullet] $, and acting on the $\hat{J}_z$ total spin operator. This series expression \eqref{qfiseries} provides the leading terms in the formulation of the QFI as an action\cite{knysh2013estimation} in eqn.\eqref{action}.

\begin{figure*}[ht!]
	\includegraphics[width=6in]{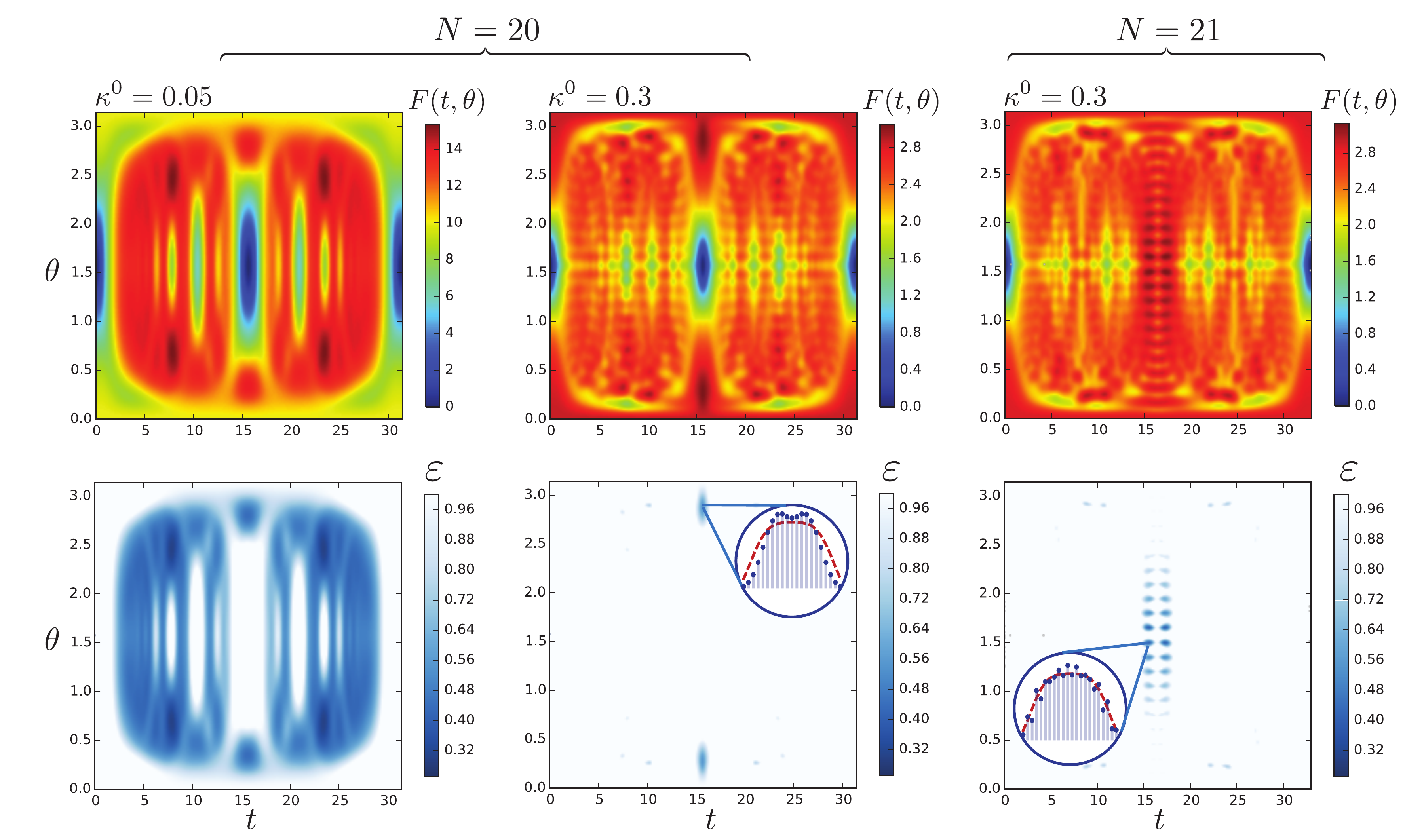}
	\caption{\label{landscape}
		\textbf{Sudden-Quench Performance}:	 The upper (rainbow) plots show the varied structure of the precision landscape for interferometric probes constructed from ($x$-direction) spin coherent states evolved by $\exp \{-i \hat{J}_y \theta \} \exp \{-i \hat{J}_z^2 t/j\}$. Quantum Fisher information $F (t,\theta)$ is also a function of decoherence. Small and large dephasing $\kappa^0$ are contrasted, and for the latter case the qualitative differences between odd and even-numbered ensembles of $N=20$ and $21$ qubits. The second row of plots indicate only those regions (blue) where the precision is supra-classical. Minimum mean squared error on any phase estimation is given by the Cramer-Rao bound: $1/F= \kappa^0 +\varepsilon/N$. For unentangled (classical) probes, $\varepsilon =1$; any smaller quantum error  $\epsilon < 1$ (blue shading) is a signature of a quantum-enhanced probe. }
\end{figure*}

\section{Decoherence function}\label{decoh-fn}
For pure collective dephasing, (background phase fluctuations with variance $\kappa^{0}$, possibly due to stochastic path length fluctuations inside the interferometer), this becomes a constant $\mu(y) \mapsto \mu_0 = N^2 \kappa^{0}$ within the physical box boundary $y = m/j \in [-1,1]$ and infinite outside the boundary. This collective dephasing is the most significant type of noise for a Bose-Einstein condensate, existing only in the fully symmetric subspace of its constituent atoms, although losses may also occur as atoms leave the condensate. Likewise, along with losses, collective dephasing is a dominant noise source in photonic interferometry. The dephasing process can be seen as a convolution of a pure probe state $|\psi\rangle$ with a Gaussian probability distribution $p_G(\theta,\bar{\theta},\kappa^0) =\exp \{-(\theta-\bar\theta)^2/ 2  \kappa^0\}/\sqrt{2 \pi \kappa^0} $ with mean $\bar{\theta}$ and variance $\kappa^0$:

\begin{equation}\label{uncert}
\rho = \int_{2 \pi}  p_G(\theta,\bar{\theta},\kappa^0) \;  | \psi (\theta) \rangle \langle \psi (\theta)  | \; \dd \theta \; .
\end{equation}
The density matrix is a mixture of these probes, each evolved by a different phase: $|\psi (\theta)\rangle = \exp \{- i \hat{J}_z \theta \}|\psi (0)\rangle$. The analysis also requires that $\kappa^0 \ll 1$. For strong phase noise beyond this limit the stochastic phase distribution can no longer be approximately Gaussian and localized within a $2 \pi$ window; periodic boundary conditions turn the random phase distribution into a wrapped normal distribution. By then, any probe state is so noisy it becomes almost completely insensitive to phase, and precision begins to decay exponentially fast in $\kappa^0$  (due to the phase uncertainty relation\cite{barnett1990quantum,knysh2014true}) . There are now two bounds on our analysis. Including the requirement that the noise parameter $\kappa^0 N^2$ is large, so that the asymptotic series can be truncated:
\begin{equation}
1/N \ll \sqrt{\kappa^0} \ll 1 \; .
\end{equation}

Collective dephasing is also important in a Bayesian estimation scheme, featuring a prior phase distribution that is updated via measurements. Dephasing is entirely equivalent to Gaussian-distributed \emph{prior} phase uncertainty\cite{macieszczak2014bayesian,personick1971application} $\Delta \theta = \sqrt{\kappa^0}$. In general, there will always be some prior phase uncertainty (estimation would be otherwise be unnecessary) and thus collective phase noise is always present

Adding local noise $ \kappa^{(L)} = \ee^\zeta -1 $ while the interferometric phase $\theta$ is being acquired is governed by a Markovian master equation: 
\begin{subequations}
	\begin{align}
	\frac{d \rho}{d \theta} &= -i [\hat{J}_z,\rho] \;  + \; \frac{d \zeta}{d \theta} \; \sum_{i=1}^N \mathcal{L}_s^{(i)} \rho \\
	\mathcal{L}_s(\rho) &= \{\hat{s}^{\dagger} \hat{s},\rho \} - 2  \hat{s} \rho \hat{s}^{\dagger}
	\end{align}
\end{subequations}
(the latter expression defines the Linbladian sueroperator $\mathcal{L}$). Noise processes are local dephasing $\hat{s} \mapsto \frac{1}{2}\hat{\sigma}_z$, excitation $\hat{s} \mapsto \frac{1}{2}\hat{\sigma}^{+}$ and relaxation $\hat{s} \mapsto \frac{1}{2}\hat{\sigma}^{-}$ defined in terms of individual Pauli spin operators. Note that $\sum_{i=1}^N \frac{1}{2} \hat{\sigma}^{(i)}_z = \hat{J}_z$. In the current analysis all qubits are permutation-invariant and decohere at the same rate; there are no topological features to this model. 
The noise function\cite{knysh2014true} then becomes:
\begin{equation}
\mu(y) = N^2 \kappa^{0} + N \kappa^{(L)} / (1-y^2). \label{hybrid-noise}
\end{equation} 
It was shown in Ref. \onlinecite{knysh2014true} that for local noise only the combination of dephasing, relaxation and excitation included in $\kappa^{(L)}$ matters asymptotically, not their individual contributions,
\begin{equation}
\zeta = \zeta_z + \zeta_{-} + \zeta_{+} \; .
\end{equation} 
In this paper, we considered the hybrid noise function $\mu(y) $ in the general form given above.

\begin{figure}
	\includegraphics[width=3in]{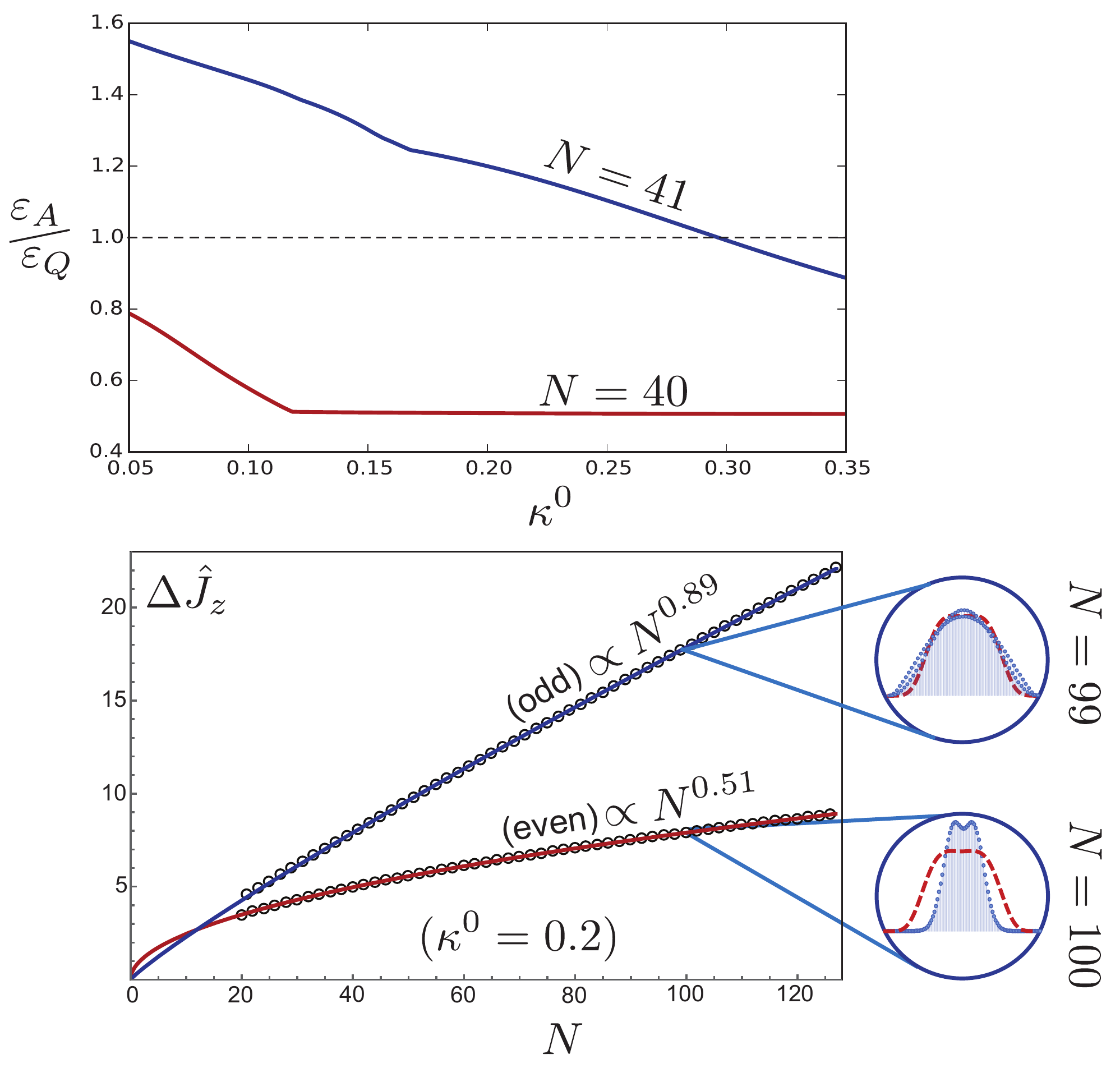}
	\caption{\label{oddeven}
		\textbf{Odd and Even Ensembles}: The associated precision of the optimal sudden-quench interferometric probe can be better/worse than the Goldilocks state in the odd/even $N$ cases respectively. For a probe subsequently subjected to collective dephasing, the upper graph indicates, as $\kappa^0$ is increased, the ratio of quantum errors  $\varepsilon_A/\varepsilon_Q$ for the optimal probes generated by annealing and a sudden quench, respectively. Below (above) the dashed horizontal line the annealed (quenched) state produces lower error. Again, the different performance of odd and even numbered $N$ ensembles is apparent. The lower illustrate some of the general characteristic differences between odd and even  sudden-quenched probes. Circle-insets show the configuration of optimal amplitudes for a sudden quench and large noise parameter. (For comparison, the optimal annealed profile, the Goldilocks probe with asymptotic width $\propto N^{2/3}$, is overlaid as a dashed red curve.)  The widths of optimal sudden-quench states are plotted as open circles and fitted to a model $\Delta \hat{J}_z = \alpha N^\beta $. The best-fit line is shown in red and blue for even and odd cases. }
\end{figure}

\section{Numerical Comparison with Sudden-Quench Dynamics}

A sidelight on our own proposal is provided by an earlier scheme\cite{ulam2001spin}, due originally to Kitagawa and Ulam-Orgikh, using entangled quantum spin states for noisy metrology. (It was analyzed recently and more comprehensively in terms of Fisher information in Ref.\onlinecite{ferrini2011effect}.) Beginning with the same spin-coherent state aligned with the strong transverse field; subsequently the field is \emph{abruptly} and discontinuously stepped to zero. The ensemble then evolves diabatically for some time under the influence of its $\sigma_z^{(1)} \sigma_z^{(2)}$ couplings. After a particular elapsed time $t$, the state is rotated by an angle $\theta$ around $\hat{J}_y$ to produce an optimal probe for noisy interferometry. The propagator is effectively $\exp \{-i \hat{J}_y \theta \} \exp \{-i \hat{J}_z^2 t/j\}$ acting on the spin coherent state. Here, for a fair comparison with our scheme, we imagine that state-preparation is decoherence-free. At least numerically, this is a straightforward optimization; even for large ensembles $N \gg1$ it involves only two bounded parameters $\theta \in [0,\pi]$ and  $t \in [0,\pi j ]$. The optimization landscape is complicated with many local minima, but in the limit of large $\kappa^0 N^2$ very few coordinate pairs $(t,\theta)$ offer supra-classical precision. FIG.\ref{landscape} depicts this landscape and optima probe shapes for $N=20$ and $21$. It is seen that precision resulting from this method can be better or worse than the Goldilocks annealed state, depending on whether the ensemble has an odd or even number of constituent spins, see FIG.\ref{oddeven}. (This may be due to the non-adiabatic unitary nature of the state preparation.)  One might now ask whether it is more feasible to prepare optimal probes by manipulating two control parameters diabatically, or just a single parameter, under the constraint that it be attenuated adiabatically. For both adiabatic and diabatic schemes the `sweet-spot' of supra-classical performance in parameter space ($\Gamma$, or $t$ and $\theta$) decreases dramatically with increasing $N$: Note the very reduced blue zone for the large noise case in the second row of FIG.\ref{landscape}.

Also, these numerics were carried out for collective phase noise, which we know favors broad probes of width $\propto N$. Most other noise types favor narrower probe distributions $\Delta \hat{J}_z \propto N^{3/4}$, according to Ref.\onlinecite{knysh2014true}. Calculation of QFI for local noise processes is a sum of terms due to state diffusion into other total spin $j = \frac{N}{2} - 1,\frac{N}{2} -2, \dots $ spaces. Necessity of numerical diagonalization in each of these additional subspaces dramatically increases computational overhead with increasing $j$, presenting a challenge for follow-up investigation.

\begin{figure}[b]
	\includegraphics[width=3.7in]{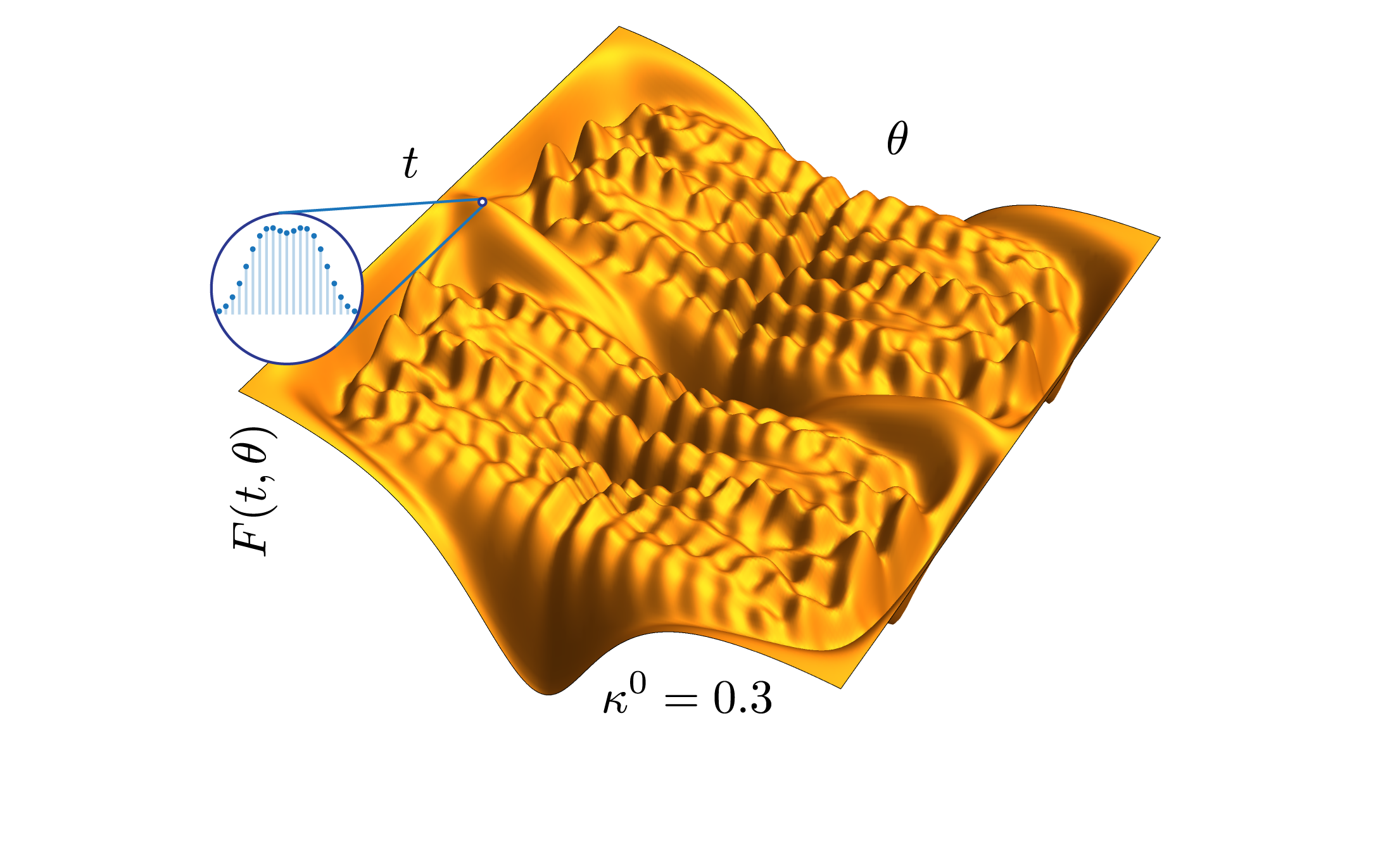}
	\caption{\label{3dlandscape}
		\textbf{Global Optimization:} For phase noise $\kappa^0 =0.3$ this figure shows the quantum Fisher information $F(t,\theta)$ for sudden-quench dynamics of $20$ qubits, representing the same data of the top middle pane of Fig.\ref{landscape}. This 3D projection better depicts the difficulty of the optimization task, illustrating the landscape's many local maxima. Optimal solutions in the space of  $t$ and $\theta$ were found by a differential evolution algorithm, which is stochastic and does not use gradient information. By comparison, a sequential least squares quadratic programming approach did not always find the global maximum,  even with multiple random starts. A basin-hopping technique or particle swarm approach might have proven effective alternatives.}
\end{figure}


\section*{Acknowledgements}
During the course of this study I had several useful discussions with Zhang Jiang and Sergey Knysh at NASA Ames Research Center. Assistance with literature review was provided by algorithms developed by the team at lateral.io

\onecolumngrid

\end{document}